\documentclass[journal]{IEEEtran}

\usepackage[utf8]{inputenc}	
\usepackage{url} 
\usepackage{siunitx} 
\usepackage{float} 
\usepackage[pdftex]{graphicx} 

\usepackage{tabularx}
\usepackage{xcolor}

\usepackage{amsmath}  
\usepackage{amssymb} 

\usepackage[noadjust]{cite} 

\newcommand{\crossvalk}[0]{4} 

\newcommand{\dissim}[0]{d} 
\newcommand{\centi}[0]{j} 
\newcommand{\filti}[0]{i} 
\newcommand{\optpatch}[0]{\mathbf{o}} 
\newcommand{\iband}[0]{b} 
\newcommand{\nband}[0]{B} 
\newcommand{\ipix}[0]{k} 
\newcommand{\npix}[0]{N} 
\newcommand{\pixpatch}[0]{\mathbb{P}} 
\newcommand{\npatch}[0]{S} 
\newcommand{\normconst}[0]{Z} 
\newcommand{\percentile}[0]{P} 
\newcommand{\covmat}[0]{\mathbf{\Sigma}} 
\newcommand{\covmathat}[0]{\mathbf{C}} 

\newcommand{\proplive}[0]{p}
\newcommand{\crownarea}[0]{A}
\newcommand{\treei}[0]{i}

\newcommand{\fte}[0]{forest-tundra ecotone}

\begin{document}

\title{Guided Nonlocal Means Estimation of Polarimetric Covariance for Canopy State Classification}

\author{Jørgen~A.~Agersborg,~\IEEEmembership{Student Member,~IEEE,}
        Stian~Normann~Anfinsen,~\IEEEmembership{Member,~IEEE,}
        and~Jane~Uhd~Jepsen,
\thanks{J.\ A.\ Agersborg and S.\ N.\ Anfinsen are with the Department of Physics and Technology, UiT The Arctic University of Norway, Tromsø, Norway. E-mail: jorgen.agersborg@uit.no.}
\thanks{J.\ U.\ Jepsen is with Norwegian Institute for Nature Research, Tromsø, Norway.}
\thanks{Manuscript received November 27, 2020; revised June 17, 2021.}} 

\maketitle

\begin{abstract}
\label{sec:abstract}
We have developed a nonlocal algorithm for estimating polarimetric synthetic aperture radar (PolSAR) covariance matrices on single-look complex (SLC) format resolution. The algorithm is inspired by recent work with guided nonlocal means (NLM) speckle filtering, where a co-registered optical image is used to aid the filtering. 
Based on patch-wise dissimilarities in the SAR and optical domains we set the weights used for the nonlocal average of the outer product of the lexicographic target vectors which form the estimate. 
By use of this method we show that the estimated covariance matrices preserve the local structure better than previous filtering methods and improve the separation of live from defoliated and dead forest. The detail preserving nature of the algorithm also means that it can be applicable in other settings where preserving the SLC format resolution is necessary. 
\end{abstract}


\section{Introduction}
\label{sec:intro}

The \fte{} is the boundary between the low-arctic tundra and the northern-boreal forest.
It is the largest vegetation transition on the planet and stretches over $\SI{13000}{\km}$ around the northern hemisphere \cite{COATscienceplan}. In places this important bioclimatic transition zone is more than 200 km wide \cite{mclaren2013boreal}.
Currently, a warming climate is contributing to both greening trends and browning trends across the circumpolar expanse of the \fte{}. Greening happens through increased vegetation productivity and/or canopy cover \cite{berner2020summer}, and browning through forest die-off caused by increased frequency of wildfires, intensified insect outbreaks and changing moisture regimes \cite{gauthier2015boreal}. Local greening processes, such as encroachment of woody species into the tundra \cite{aune2011contrasting, rees2020subarctic, myers2020complexity}, may be counteracted by herbivores such as browsing ungulates or defoliating forest pest insects.

The \fte{} in northern Fennoscandia is dominated by mountain birch (Betula pubescens var. pumila L.). In this system, decadal outbreaks by several species of geometrid moth cause defoliation and tree mortality. The outbreak ranges of several of these moth species have expanded in recent decades, causing intensified multi-species outbreaks \cite{jepsen2011rapid}. These outbreaks have been shown to lead to rapid transitions of the state of the \fte{} \cite{vindstad2018pest}. 
Defoliating forest pest species, such as geometrids, usually do not kill their host tree outright, but inflict damage that when accumulated over several years, and often in combination with other stressors, can exceed the resilience of the host tree and lead to extensive tree mortality \cite{jepsen2013ecosystem, senf2017remote, tenow1972outbreaks}. 
The outbreaks may be spatially extensive and cover hundreds of $\si{\km^2}$. However, since the degree of tree mortality and the recovery of the crown layer are influenced by both biotic and abiotic factors, including ungulate browsing, exposure, soil moisture and soil quality, an outbreak typically leaves a complex mosaic of forest stands with both sharp and fuzzy boundaries between recovering, partly recovering and dead stands. 

Remote sensing from satellite provides a valuable contribution to the identification and quantification of these processes by being able to monitor the effects of geometrid outbreaks on canopy state and regrowth for vast areas. The approach taken in previous work is to detect proxies of canopy defoliation based on coarse resolution (pixel resolution $> \SI{200}{\metre}$) normalised difference vegetation index (NDVI) products derived from multispectral optical remote sensing images, and correlating this with field work measurements of larvae densities \cite{jepsen2009monitoring, olsson2016near, olsson2016development}. This follows the convention that defoliation studies often are based on NDVI products. A literature review published in 2017 shows that 82 \% of studies mapping defoliation of broadleaved forest caused by insect disturbance used a single spectral index, and most frequently NDVI \cite{senf2017remote}.

In this work, we will consider synthetic aperture radar (SAR) as a data source for primarily three reasons. Firstly, polarimetric SAR data is theoretically able to differentiate between scattering mechanisms such as surface, volume, and double bounce. Hence the information contained in the polarimetric covariance matrices could be able to separate live canopy state (volume scattering from leaves) from defoliated and dead canopy states (double bounce scattering mainly from the trunks). Secondly, remote sensing products from satellite-based SAR are near weather-independent. High latitude areas, such as subarctic Fennoscandia, have a high average cloud cover percentage, which limits observations by optical satellites. Thirdly, it is of interest to evaluate how SAR performs when it comes to monitoring canopy state. While SAR has been used to monitor \emph{deforestation}, none of the studies of broadleaved forest \emph{defoliation} summarised in \cite{senf2017remote} used SAR.
Also, SAR has been used for several studies for characterising tundra environments and their dynamics, see \cite{ullmann2017scattering} and references therein. It has also been used for monitoring tundra shrub growth and expansion \cite{duguay2015potential}.

For remote sensing to contribute to understanding the complicated dynamics of the \fte{}, it is important to be able to separate between different canopy states, most noticeably intact leaf cover versus missing or significantly reduced leaf cover. 
While the difference in scattering mechanisms may in theory be able to separate the two classes, the inherent speckle phenomenon of coherent imaging systems such as SAR means that filtering is necessary before the data can be utilised.
For polarimetric SAR data the speckle filtering is often performed in connection with the estimation of the covariance matrices. 
However, traditional speckle filtering comes at the cost of spatial resolution.
To obtain the polarimetric covariance matrices and still retain as much spatial resolution as possible to better classify the sparse interwoven forest of the \fte{}, we have developed a new estimation algorithm, named polarimetric guided nonloacal means (PGNLM).
It is inspired by the guided nonlocal means (GNLM) speckle filtering algorithm which employed an optical guide image to filter intensity SAR data in \cite{vitale2019gnlm}. 
We show that PGNLM gives an improvement in the estimation of the covariance matrices compared to standard methods, especially when it comes to preserving details. We also demonstrate that polarimetric covariance matrix data estimated with the PGNLM algorithm better separate between live and dead forest in our area of interest (AOI).

The rest of the paper is organised as follows. Section \ref{sec:pgnlm} begins by describing the GNLM algorithm, before introducing our method. Other related work is summarised in Section \ref{sec:related}, while the data from out area of interest and the satellite imagery are presented in Section \ref{sec:method}. The results, both for comparison of PGNLM with traditional filtering methods and for the canopy state classification, are given in Section \ref{sec:results}. Section \ref{sec:conclusion} concludes and summarises the work.

\section{Polarimetric Guided Nonlocal Means}
\label{sec:pgnlm}

\subsection{Nonlocal means filtering}
\label{subsec:nonlocal-means}
The nonlocal means (NLM) image filtering algorithm was first introduced by Buades \emph{et al.\ }in \cite{buades2005review}.
Nonlocal algorithms are based on splitting the denoising problem into two steps: 1) Finding good predictors; and 2) using these predictors in the estimation \cite{deledalle2014exploiting}.
Whereas boxcar-type algorithms make the implicit assumption that the closest pixels make the best predictors for estimating the "clean signal", nonlocal algorithms use a similarity criterion to find possibly disjoint sample points in a wider search area.
Many different similarity measures, both based on individual pixels and patches with a width and height of several pixels, have been used to find and weight contributions to the filtered output image. 
The use of a guide image to aid the filtering of a noisy input image was first published by He \emph{et al.\ }in \cite{he2012guided}. The guide image could be the noisy input itself, a pre-filtered version of it, or another coregistered image \cite{he2012guided}.

\subsection{Nonlocal speckle filtering with optical guide image}
\label{subsec:optical-guide-sar-filtering}

The use of a coregistered optical image to guide the SAR despeckling was first proposed in \cite{verdoliva2014optical-driven-nonlocal}.
An important aspect of the methodology is that the filtered output is a combination of SAR pixels only, to avoid injecting optical image geometry into the SAR scene \cite{verdoliva2014optical-driven-nonlocal}.
Gaetano \emph{et al.\ }\cite{gaetano2017fuse-sar-opt} extended this work to use a nonlocal means framework, except in strongly heterogeneous areas of the SAR scene, since areas with strong point scatterers and coherent scattering are an inherent part of radar images that should be preserved when filtering.
Also \cite{gaetano2017fuse-sar-opt} extended the previous work by switching to patch-based filtering, where image patches rather and individual pixels were averaged. 
The need to explicitly test for heterogeneous areas was replaced by reliability tests that remove unreliable predictors as a further step of development \cite{vitale2019gnlm}.

The previously mentioned work covers single-channel intensity SAR images only \cite{verdoliva2014optical-driven-nonlocal, gaetano2017fuse-sar-opt, vitale2019gnlm}. The filtering problem can in this case be formulated as trying to estimate the "clean" intensity image $\mathbf{\hat{X}}$ based on the original noisy intensity data $\mathbf{X}$, and with the help of the coregistered optical guide image $\mathbf{O}$. The filtering is patch-based, where a patch centred on a pixel with spatial index (pixel position) $\centi$ is defined as $\mathbf{x}(\centi) = \{ \mathbf{X}(\centi + \ipix) \, , \, \ipix \in \pixpatch \}$, where $\pixpatch$ indicates a set of $\npix$ spatial offsets with respect to $\centi$ \cite{vitale2019gnlm}.
The filtering is then done for each patch $\mathbf{x}(\centi)$ centred on pixel $\centi$ in the input SAR image, by summing the weighted patches $\mathbf{x}(\filti)$ in a search area $\Omega(\centi)$ around $\centi$:
\begin{equation}
  \label{eq:gnlm-pix}
  \mathbf{\hat{x}}(\centi) = \sum\limits_{\filti \in \Omega(\centi)} w(\filti,\centi) \mathbf{x}(\filti)
\end{equation}
where  the size of the search area $\Omega$ is determined by a parameter, and the patch size of $\mathbf{\hat{x}}$ and $\mathbf{x}$ are equal and given by $\pixpatch$. Since each pixel is part of multiple patches, the filtering procedure will estimate each pixel multiple times \cite{vitale2019gnlm}.

Note that the optical data does not enter into Eq.\ \eqref{eq:gnlm-pix} directly, and only SAR domain pixels are used in the filtering \cite{vitale2019gnlm}. The guide is only used to help set the weights $w(\filti,\centi)$ used in Eq.\ \eqref{eq:gnlm-pix}.
The optical guide image should have the same pixel size as the SAR image and must be coregistered to it. 
The weights are based on $\dissim_\text{SAR}$ and $\dissim_\text{OPT}$, which are the patch-based dissimilarity measures in the SAR and optical domain, respectively.
The dissimilarities are mapped into weights by the exponential kernel, 
so the weight determining how much the filtering of a patch centred on pixel $\centi$ is influenced by a patch centred on pixel $\filti$ can be expressed as: 
\begin{equation}
\label{eq:weight-org}
    w(\filti,\centi) = \normconst e^{ - \lambda \left[ \gamma \dissim_\text{SAR}(\filti,\centi) + (1-\gamma) \dissim_\text{OPT}(\filti,\centi) \right]}
\end{equation}
where $\normconst$ is a normalising constant, $\lambda$ is an empirical weight parameter, and $\gamma \in \left[ 0, 1 \right]$ balances the emphasis on SAR versus optical dissimilarity.

For the optical domain dissimilarity $\dissim_\text{OPT}$, \cite{vitale2019gnlm} used the normalised sum of squared Euclidean distances:
\begin{equation}
\label{eq:opt-dissim}
    \dissim_\text{OPT}(\filti,\centi) = \frac{1}{\nband \npix} \sum_{\iband=1}^\nband \sum_{\ipix \in \pixpatch} \left[ \optpatch_{\iband , \filti + \ipix} - \optpatch_{\iband, \centi + \ipix} \right]^2
\end{equation}
where $\nband$ is the number of bands in the optical guide and $\npix$ is the number of pixels in each patch, determined by the set of spatial offsets $\pixpatch$. The SAR dissimilarity $\dissim_\text{SAR}$ was defined as 
\begin{equation}
\label{eq:gnlm-sar-dissim}
    \dissim_\text{SAR}(\filti,\centi) = \frac{1}{\mu_D \npix} \sum_{\ipix \in \pixpatch} \log \left[  \frac{ \mathbf{x}_{\filti + \ipix} + \mathbf{x}_{\centi + \ipix} }{ 2 \sqrt{\mathbf{x}_{\filti + \ipix} \mathbf{x}_{\centi + \ipix} } } \right]
\end{equation}
with $\npix$ defined as before, and where $\mu_D$ is a factor based on the statistical assumptions made, which should ensure that in strictly homogeneous regions the distance between target and predictor patch is unitary \cite{vitale2019gnlm}. 

While the weights in Eq.\ \eqref{eq:weight-org} should de-emphasise dissimilar patches, two steps are taken to discard entirely a number of "unreliable predictors" \cite{vitale2019gnlm}.
First, patches with too high SAR dissimilarity with the patch being estimated are excluded. This is especially important to prevent high intensity pixels from being reproduced in the patch being estimated; despite the patches being dissimilar, the low weight would be offset by the strong signal \cite{vitale2019gnlm}. Thus, the set of patches used for averaging in Eq.\ \eqref{eq:gnlm-pix}, $\Omega(\centi)$, is replaced by a new set $\Omega'(\centi)$ where 
\begin{equation}
\label{eq:omega-1prime}
    \Omega'(\centi) = \{ \filti \in \Omega(\centi) : \dissim_\text{SAR}(\filti,\centi) < T_\text{SAR} \}
\end{equation}
and $T_\text{SAR}$ is the threshold which determines for which dissimilarity level patches are ignored. In the extreme case where no other patches satisfy this threshold, perhaps due to corner reflectors in the patch being estimated, that patch is not filtered \cite{vitale2019gnlm}. However, individual pixels may be filtered due to being member of many patches. Note that the self-dissimilarity of a patch is always zero, as can easily be seen from Eq.\ \eqref{eq:opt-dissim} and \eqref{eq:gnlm-sar-dissim}, so the patch being estimated, centred on $\centi$, will always be in $\Omega'(\centi)$. 

In addition to the step in Eq.\ \eqref{eq:omega-1prime}, a second modification to the set of patches was used \cite{vitale2019gnlm}. To prevent oversmoothing, the maximum number of patches that could be used for averaging was reduced from the total number of patches in the full search area size, $\npatch$, to $\npatch_0$ with $\npatch_0 < \npatch$, even if all predictors met the criterion in Eq.\ \eqref{eq:omega-1prime}. The new set, $\Omega''(\centi) \subseteq \Omega'(\centi)$, is chosen from those patches of $\Omega'(\centi)$ with the smallest optical domain dissimilarity, $\dissim_\text{OPT}$ \cite{vitale2019gnlm}. The new set has the cardinality
\begin{equation}
\label{eq:omega-2prime-card}
    |\Omega''(\centi)| = \min \{|\Omega'(\centi)|, \npatch_0 \}
\end{equation}
Thus, the final filtering result is based on Eq.\ \eqref{eq:gnlm-pix}, but with the set of patches being modified by the thresholding in Eq.\ \eqref{eq:omega-1prime}, and with the final number of patches used for averaging limited by Eq.\ \eqref{eq:omega-2prime-card}. The weights are given by Eq.\ \eqref{eq:weight-org}. According to the terminology used by Hu \emph{et al.\ }\cite{hu2015non}, the filtering is a combination of an exponential kernel and what they call a "piecewise kernel" resulting from the thresholding operation.

\subsection{Polarimetric covariance matrix estimation}
\label{subsec:general-covmat-estimation}

The multilook complex covariance matrix, $\covmat$, is commonly used as a basis for characterising the different scattering phenomena when working with polarimetric SAR data.
For each pixel, the complex lexicographic target vector can be written as,
\begin{equation}
  \label{eq:s-vec}
  \mathbf{s} = \left[ S_{{\text{HH}}}, S_{{\text{HV}}}, S_{{\text{VV}}} \right]^T \in \mathbb{C}^{3 \times 1}\,,
\end{equation}
where the first subscript indicates the polarisation of the transmitted pulse and the second the received polarisation (horizontal (H) or vertical (V)). We have assumed reciprocity, $S_{\text{HV}} = S_{\text{VH}}$, such that the scattering coefficients of the cross-polarisation channels (HV and VH) have already been averaged and stored as $S_{\text{HV}}$. Given a set of target vectors $\{\mathbf{s}_i\}_{i=1}^{n}$ an estimate of $\covmat$, the sample covariance matrix $\covmathat$, can be computed as the sample mean
\begin{equation}
  \label{eq:c3-mat}
  \covmathat = \langle \mathbf{s} \mathbf{s}^H \rangle = \frac{1}{n}\displaystyle\sum_{i=1}^{n}\mathbf{s}_i\mathbf{s}_i^H
\end{equation}
where the brackets $\langle\cdot\rangle$ denote averaging over a data sample and the superscript $H$ the conjugate transpose operation. 
A speckle filtering step can be combined with the calculation of $\covmathat$, where the different speckle filtering algorithms determine which complex scattering vectors are used in the averaging operation in Eq.\ \eqref{eq:c3-mat}, or in a similar weighted average.
While speckle should be suppressed, the polarimetric information should also be preserved \cite{xing2017feature}.
The general idea is to avoid mixing pixels belonging to different structures, so that edges and strong scatterers are preserved by not averaging such pixels with their heterogeneous background or adjacent areas \cite{deledalle2014nl}.  

Another common way of representing the polarimetric information is by the coherency matrix $\mathbf{T}$. However, since $\covmathat$ and $\mathbf{T}$ are linked by a unitary transformation, we will in the following only look at the former. 
Also, when referring to other work we will use the term covariance matrix for notational consistency, even if the method is based on the coherency matrix.

\subsection{Nonlocal polarimetric covariance matrix estimation}
\label{subsec:nonlocal-covmat-estimation}
\label{subsec:discarding-predictors} 
\label{subsec:normalising-dissims} 

To use a GNLM framework to estimate polarimetric covariance matrices, two adaptations to the method need to be made. The first is to exchange the SAR dissimilarity measure in Eq.\ \eqref{eq:gnlm-sar-dissim} with one where the polarimetric information is taken into account. The second and more extensive adaptation is to modify the patch-based filtering of intensity pixels in Eq.\ \eqref{eq:gnlm-pix} to estimate the covariance matrix $\mathbf{C}$. 
In addition, we have modified how unreliable predictors are discarded and how the parameters of the algorithm are set.

The SAR dissimilarity measure used in \cite{vitale2019gnlm} is for single polarisation intensity data, and the values in the numerator and denominator inside the logarithm of Eq.\ \eqref{eq:gnlm-sar-dissim} are scalars. 

We need to define a dissimilarity measure for PolSAR data. 
One approach to defining a (dis)similarity measure could be to base it on classical hypothesis tests from multivariate statistics. 
An incoherent target vector $\mathbf{s}$ representing natural terrain is commonly assumed to follow a multivariate complex circular Gaussian distribution with expectation equal to the zero vector, $\mathbb{E}\{\mathbf{s}\} = \mathbf{0}$, and a covariance matrix $\covmat$ which contains all the information that characterises the scene \cite{torres2014speckle}.
Since the target vectors are assumed to be zero mean, a test that compares multivariate means between two samples is not an alternative since they would be equal. 
On the other hand, several works (\cite{chen2010nonlocal, zhong2013robust, torres2014speckle, deledalle2014nl}) have used (dis)similarity measures for the covariance matrices based on the complex Wishart model. 
However, using such test statistics requires that we first estimate the covariance matrices locally. In the works mentioned above, a covariance-based test statistic is used in the (dis)similarity measure, which again is used to set the weights for nonlocal averaging of the initial covariance matrix estimate.

Here we seek to avoid the local pre-estimation of covariance matrices and the implicit assumption that pixels which are spatially close have the same scattering properties.
This is motivated by the sparse nature and the fine-scale variability of the forest in the \fte{}. 
Instead we want to weight the terms in Eq.\ \eqref{eq:c3-mat} according to the similarity with the pixel position to be estimated, and only average over outer products which are sufficiently similar. 
The goal of avoiding local estimation of covariance matrices altogether implies that we must base our dissimilarity measure on the complex target vectors $\mathbf{s}$.
Since methods based on testing differences between mean vectors are not feasible, we use a heuristic dissimilarity measure based on the Euclidean vector norms of the target vectors.
More specifically, we use the norm of the difference between two target vectors and divide by the average norm of the two target vectors.
We rely on the patch-based nature of the dissimilarity and the optical guide to find good weights. Furthermore, we use two criteria based on Eqs.\ \eqref{eq:omega-1prime} and \eqref{eq:omega-2prime-card} for discarding bad predictors. 
Our covariance matrix estimate will therefore avoid any local pre-estimation and will work directly on the PolSAR data in SLC format. 
 
We first start by defining a dissimilarity measure between two target vectors.
A measure of the dissimilarity $\dissim$ between target vectors at position $\centi$ and $\filti$ can be defined as
\begin{equation}
  \label{eq:svec-dissim}
  \dissim( \mathbf{s}_\centi, \mathbf{s}_\filti) = \frac{(\mathbf{s}_\centi - \mathbf{s}_\filti)^H (\mathbf{s}_\centi - \mathbf{s}_\filti) }{ \frac{1}{2} ( \mathbf{s}_\centi^H \mathbf{s}_\centi +  \mathbf{s}_\filti^H \mathbf{s}_\filti )}.
\end{equation}
We note that Eq.\ \eqref{eq:svec-dissim} also can be written as
\begin{equation}
  \label{eq:svec-norm-dissim}
  \dissim( \mathbf{s}_\centi, \mathbf{s}_\filti) = \frac{||\mathbf{s}_\centi - \mathbf{s}_\filti)||_2^2 }{ \frac{1}{2} ( ||\mathbf{s}_\centi||_2^2 +  ||\mathbf{s}_\filti||_2^2 )} 
\end{equation}
where $||\cdot||_2^2$ denotes the squared Euclidean norm of the complex vectors. 
We want $\dissim( \mathbf{s}_\centi, \mathbf{s}_\filti)$ to measure the dissimilarity in scattering types of the two target vectors.
The denominator in Eq.\ \eqref{eq:svec-dissim} is necessary to avoid that the strength of the received signal (backscatter) heavily influences the dissimilarity. 
This is especially important when comparing two target vectors with low backscatter, where the dissimilarity would be low if it was only based on the numerator of Eq.\ \eqref{eq:svec-norm-dissim}, even if the relative difference between $\mathbf{s}_\centi$ and $\mathbf{s}_\filti$ was significant due to different scattering phenomena.
We have opted to use the average of the squared norms in the denominator, rather than the product of the norms, as this gives better numerical behaviour of the dissimilarity measure when one of the vector norms are close to zero.
In our previous work \cite{myIGARSS2020} we did not include the norm of the patch centred on the candidate target vector, $||\mathbf{s}_{\filti}||_2^2$, in the denominator of Eq.\ \eqref{eq:svec-dissim}. 
Without the second norm in the denominator, the dissimilarity between a target vector from a low backscatter region and a high backscatter region would depend on which of the two the search area was centred on, i.e.\ which of them was used in the denominator. 
Using both norms makes the dissimilarity between two vectors symmetric.

To obtain a patch-based dissimilarity between the patches centred on pixel position $\centi$ and pixel position $\filti$, we sum the expression in Eq.\ \eqref{eq:svec-dissim}: 
\begin{equation}
  \label{eq:new-sar-dissim}
  \dissim_\text{Pol}(\filti,\centi) = \frac{1}{\npix} \sum_{\ipix \in \pixpatch} \frac{  (\mathbf{s}_{\centi +\ipix} - \mathbf{s}_{\filti +\ipix})^H (\mathbf{s}_{\centi +\ipix} - \mathbf{s}_{\filti +\ipix})  }{ \frac{1}{2} ( \mathbf{s}_{\centi +\ipix}^H \mathbf{s}_{\centi +\ipix}  +  \mathbf{s}_{\filti +\ipix}^H \mathbf{s}_{\filti +\ipix})  }
\end{equation}
where $\npix$ and $\pixpatch$ are defined as for Eqs.\ \eqref{eq:opt-dissim} and \eqref{eq:gnlm-sar-dissim}. 
The exact numerical value of the dissimilarity is not so important due to a normalisation before its use in the weight function.
Eq.\ \eqref{eq:new-sar-dissim} could also be written in terms of the squared Euclidean norm, just as in Eq.\ \eqref{eq:svec-norm-dissim}:
\begin{equation}
  \label{eq:polsar-norm-dissim}
  \dissim_\text{Pol}(\filti,\centi) = \frac{1}{\npix} \sum_{\ipix \in \pixpatch} \frac{  ||\mathbf{s}_{\centi +\ipix} - \mathbf{s}_{\filti +\ipix})||_2^2  }{ \frac{1}{2} ( ||\mathbf{s}_{\centi +\ipix}||_2^2  +  ||\mathbf{s}_{\filti +\ipix}||_2^2)  } .
\end{equation}
The patch-based nature of the dissimilarity measure means that the patches compared should be structurally similar.
This makes the dissimilarity measure more robust. It is particularly beneficial when the patch contains a detail such as an edge, since patches without the same edge will get a higher value of $\dissim_\text{Pol}$.
The ability to match structurally similar patches may not be as relevant for natural terrain as for areas with a prevalence of man-made structures. Despite the potential rejection of patches due to structural dissimilarity, we expect to find a high number of relevant predictors also in natural terrain, especially since we anticipate high pixel similarity in the optical guide for such areas.

The second major adaptation of the GNLM method is to go from intensity image speckle filtering to polarimetric covariance matrix estimation.
To achieve this we introduce a weighting based on the PolSAR and optical guide dissimilarities to Eq.\ \eqref{eq:c3-mat}.
The estimate will then have the same form as Eq.\ \eqref{eq:gnlm-pix}, and we can then find the polarimetric guided nonlocal means (PGNLM) estimate for the covariance matrix as:
\begin{equation}
    \label{eq:pgnlm-c3}
    \mathbf{C}(\centi) = \frac{1}{\normconst(\centi)} \sum\limits_{\filti \in \Omega''(\centi)} w(\filti,\centi) \mathbf{s}_\filti \mathbf{s}_\filti^H
\end{equation}
where $\normconst(\centi)$ is a normalisation factor and the weight $w(\filti,\centi)$ is defined similarly to Eq.\ \eqref{eq:weight-org}, but with some modifications such as $\dissim_\text{SAR}$ being replaced by a PolSAR dissimilarity measure and the normalisation factor being included in the denominator of Eq.\ \eqref{eq:pgnlm-c3} instead. The final expression for the PGNLM weight is defined later in Eq.\ \eqref{eq:weight-pgnlm}. The normalisation factor is the sum of the weights for the covariance matrix estimate for each pixel position:
\begin{equation}
    \label{eq:pgnlm-normconst}
    \normconst(\centi) = \sum\limits_{\filti \in \Omega''(\centi)} w(\filti,\centi) . 
\end{equation}

The use of patch-based dissimilarities for calculating the weights means that the outer products, $\mathbf{s}_\filti \mathbf{s}_\filti^H$, averaged to form the covariance matrix come from similar regions, which helps prevent spurious matches that could occur if the weights were based on the dissimilarity between two target vectors, such as using Eq.\ \eqref{eq:svec-dissim} directly. 
Note that in the speckle filtering of single-channel intensity in Eq.\ \eqref{eq:gnlm-pix}, each pixel intensity is estimated multiple times, as it is a part of multiple patches.
This also means that it is more convenient to include the normalisation factor in Eq.\ \eqref{eq:weight-org} than in Eq.\ \eqref{eq:gnlm-pix}.
Contrarily, the covariance matrix for pixel position $\centi$ is only estimated once. It is the weighted sum of outer products from pixel positions where the patch centred on that pixel is sufficiently similar to the patch centred on $\centi$.
Thus, the patches $\pixpatch$ should preferably have sides with an odd number of pixels, since the centre of the patch then corresponds to a pixel centre position.
As the target vectors averaged in Eq.\ \eqref{eq:pgnlm-c3} represent pixels and not patches, unlike the intensities averaged in Eq.\ \eqref{eq:gnlm-pix}, we shall refer to the outer products in the weighted average of Eq.\ \eqref{eq:pgnlm-c3} as predictors to maintain a clear terminology.

For GNLM it was found that the algorithm relied heavily on the "reliability test’s capacity of rejecting bad predictors" \cite{vitale2019gnlm}. 
These rejection criteria are summarised in Equations \eqref{eq:omega-1prime} and \eqref{eq:omega-2prime-card}. 
In the PGNLM estimate of polarimetric covariance matrices we also want to discard unreliable predictors. 
We perform the same two steps as in \cite{vitale2019gnlm}: first we set an upper limit to the SAR domain dissimilarity with a threshold to obtain a modified set of predictors, $\Omega'(\centi)$, according to Eq.\ \eqref{eq:omega-1prime}; 
then we set a maximum number of predictors $\npatch_0 < \npatch$ that can be used, where the predictors are limited by discarding the candidates with the highest optical domain dissimilarity, thus obtaining the final set $\Omega''(\centi)$ from Eq.\ \eqref{eq:omega-2prime-card}.

Vitale \emph{et al.\ }\cite{vitale2019gnlm} included an analysis of the choice of threshold $T_\text{SAR}$. As a lower baseline it was set to $T_\text{SAR} = 1$, called a "very small threshold", and the effect of adding various multiples of the standard deviation $\sigma$ "of the normalised SAR distance for homogeneous signal" \cite{vitale2019gnlm} was studied.
Rather than performing a similar analysis for our dissimilarity measure, we choose a different approach that avoids making assumptions about the data distribution and seeks to keep the threshold parameter sensor-independent. 
Henceforth, we will refer to this threshold on $\dissim_\text{Pol}$ as $T_\text{Pol}$. Furthermore, instead of requiring the user to set the threshold value $T_\text{Pol}$ directly, the user sets a percentile $\percentile_\text{Pol}$ that, based on a pre-calculated reference set of typical dissimilarities for the image, determines $T_\text{Pol}$.
A motivation for this is to facilitate that the parameters can be set without manual analysis of the data in advance.
By setting the SAR dissimilarity threshold to a percentile, we have also limited the search space for this parameter, since the value of $\percentile_\text{Pol}$ is between $0$ and $100$.
We shall now examine how the reference set of dissimilarities can be calculated.

\begin{figure}[htb]
  \centering
  \includegraphics[width=1.0\linewidth, keepaspectratio]{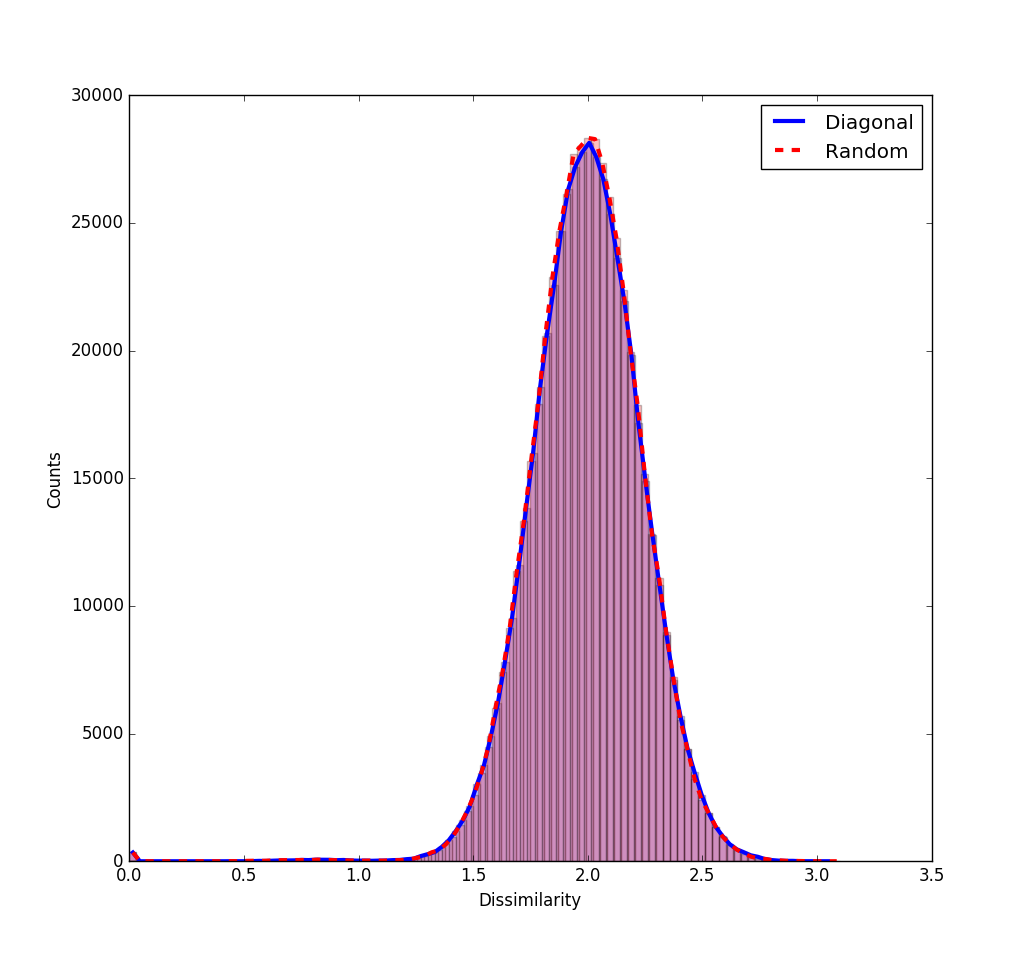}
\caption{Histogram of PolSAR dissimilarities $\dissim_\text{Pol}$.}
\label{fig:sar_dissim_hists}
\end{figure}

Fig.\ \ref{fig:sar_dissim_hists} shows histograms of two sets of patch-wise dissimilarities that are computed in different ways. One set is produced by centring the search area of $39 \times 39$ pixels along the main diagonal of a $384 \times 524$ test dataset and calculating patch-wise dissimilarities between the $5\times 5$ centre patch and all possible $5\times 5$ patches within the search area. This gives a total of $520,182$ dissimilarity values\footnote{ 
The main diagonal contains $384$ elements. With a search area of $39 \times 39$, the distance from the centre pixel to the corner of the square is 19 pixels. Furthermore, only the centre of a patch needs to be within the search area, and with $5\times 5$ patches this means that an extra 2 pixels is needed to be within the valid region of the image. Thus, the search area needs to be placed $19+2=21$ pixels into the image. Since this must be done for both ends of the main diagonal, the total number of valid search areas are $384-2*21=342$. 
Each search area gives $39 \times 39 = 1521$ dissimilarity values. 
Thus the total number of dissimilarity values are $342*1521=520,182$.
Note that unlike the threshold calculation, both the PolSAR data and the optical guide are padded when calculating the weights. }. The other set is computed by centring the search areas on image coordinates that are randomly drawn (without replacement) and calculating the same amount of dissimilarities between centre patches and displaced patches in the search area. 
Fig.\ \ref{fig:sar_dissim_hists} shows the resulting dissimilarity histograms for the two sets.
Since there is no apparent difference between using a set of randomly positioned search areas and using search areas along the diagonal, the experiments in this paper will use the latter method for computing $\percentile_\text{Pol}$ to enable a reproducible and stable comparison between the different parameter settings.  

When estimating the SAR dissimilarities $\dissim_\text{Pol}$ to obtain the threshold $T_\text{Pol}$ corresponding to the selected $\percentile_\text{Pol}$ percentile, we also estimate the optical dissimilarity $\dissim_\text{OPT}$ for the same patches using Eq.\ \eqref{eq:opt-dissim}.
Likewise, we also specify a percentile $\percentile_\text{OPT}$ for the optical dissimilarities and use it to obtain a corresponding percentile value or threshold $T_\text{OPT}$ for the optical dissimilarities.
We then divide both dissimilarities with their respective percentile values
\begin{align}
  \Tilde{\dissim}_\text{Pol} & = \dissim_\text{Pol} / T_\text{Pol} \label{eq:polsar-tilde-dissim} \\
  \Tilde{\dissim}_\text{OPT} & = \dissim_\text{OPT} / T_\text{OPT} \label{eq:opt-tilde-dissim}
\end{align}
where $T_\text{Pol}$ and $T_\text{OPT}$ are the estimated percentile values of $\dissim_\text{Pol}$ and $\dissim_\text{OPT}$ for the $\percentile_\text{Pol}$ and $\percentile_\text{OPT}$ percentiles, respectively.
We use these "normalised" dissimilarity values for updating the exponential weight function in Eq.\ \eqref{eq:weight-org}.
The weights for the PGNLM algorithm are then defines as
\begin{equation}
\label{eq:weight-pgnlm}
    w(\filti,\centi) = \exp \left[  - \lambda ( \gamma \Tilde{\dissim}_\text{Pol}(\filti,\centi) + (1-\gamma) \Tilde{\dissim}_\text{OPT}(\filti,\centi) ) \right] 
\end{equation}
where the kernel scale $\lambda$ is an empirical weight parameter, and $\gamma \in \left[ 0, 1 \right]$ is the emphasis on PolSAR versus optical dissimilarity.

Since any candidate predictor where the PolSAR patch dissimilarity exceeds the threshold is discarded, we know that the normalised dissimilarity values are between zero and one, so $\Tilde{\dissim}_\text{Pol} \in [0,1]$. 
The same does not hold true for $\Tilde{\dissim}_\text{OPT}$, as predictors are not discarded based on the value directly, but rather the number of predictors are capped at $\npatch_0$ based on the lowest values of the optical dissimilarity. Dividing by a percentile value means we have more control over the range of $\Tilde{\dissim}_\text{OPT}$. If we set the percentile $\percentile_\text{OPT}$ higher than the fraction of predictors used, $\npatch_0 / \npatch$, it is probable that the highest optical dissimilarity will not exceed $1$, since these predictors are unlikely to be included in the set of predictors $\Omega''(\centi)$. 
The lowest possible weight will depend on the kernel scale parameter $\lambda$, as well as the percentiles $\percentile_\text{Pol}$ and $\percentile_\text{OPT}$.
The higher the percentiles, the more weight will be assigned to the predictors. For the kernel scale $\lambda$ the opposite is true: the higher the value, the less weight will be assigned to the predictors.
While the lowest possible weight is not strictly bounded, a predictor candidate that has a dissimilarity in the SAR and optical domains that are equal to the percentile values, $\dissim_\text{Pol} = T_\text{Pol}$ and $\dissim_\text{OPT} = T_\text{OPT}$, will have a weight of $e^{-\lambda}$ since $\Tilde{\dissim}_\text{Pol} = 1$ and $\Tilde{\dissim}_\text{OPT} = 1$.
It is easy to see from Eqs.\ \eqref{eq:opt-dissim} and \eqref{eq:new-sar-dissim} that the lowest possible dissimilarity is the self-dissimilarity of a patch with itself, which is $0$. 
This corresponds to a weight of $w(\centi,\centi) = 1$, which is the highest possible weight.
Hence, the kernel scale parameter $\lambda$ can and should be set with these two cases in mind. Specifically, since the highest possible weight is $1$, the value of $e^{-\lambda}$ can be considered as the relative weight fraction that is assigned to a predictor whose patchwise PolSAR and optical dissimilarities both are right on the user specified percentile tolerance limits.
For the balancing parameter $\gamma$, the normalisation of $\Tilde{\dissim}_\text{Pol}$ and $\Tilde{\dissim}_\text{OPT}$ means that both dissimilarities are in approximately the same range, and that range is known.
Thus, $\gamma$ does not need to consider different ranges of the dissimilarity values for the two domains, and can be used to emphasise which of them should contribute most to the weight.   

The normalisation of the dissimilarities makes the parameter selection easier, especially when working with new datasets, but at the expense of computation time.
Thus, ad hoc adjustments to the parameters when using data from different sensors can be avoided.

Our Python implementation of the algorithm is provided in GitHub: \url{https://github.com/jorag/pgnlm/}.  
Included in the repository is a simulated toy PolSAR dataset with a coregistered optical guide.

\section{Related work}
\label{sec:related}

Apart from the GNLM algorithm and general nonlocal image filtering, which are summarised in Section \ref{sec:pgnlm}, other related works are discussed in this section.
Many different nonlocal algorithms have been used for speckle filtering since nonlocal means image filtering was first introduced in \cite{buades2005review}. Often the similarity criteria are based on patch-wise similarity measures for robustness. See \cite{deledalle2014exploiting} and references therein for an overview. 
However, nonlocal algorithms for \emph{filtering} of covariance matrices traditionally estimate the covariance matrices locally, and then filter them nonlocally. We, on the other hand, \emph{estimate} the covariance matrices in the nonlocal algorithm, and utilise this to maintain SLC resolution for the resulting covariance matrices. In a local approach this could only be achieved by using a sliding window in the sample mean estimate, which would result in an interpolation that suffers from severe smoothing and loss of detail. 

An early example of a nonlocal speckle filter for polarimetric covariance matrices is \cite{chen2010nonlocal}, where a test statistic is used as a "pretest" to select similar pixels. The likelihood ratio test statistic is based on the assumption that the covariance matrices have a complex Wishart distribution \cite{conradsen2003test}, thus they have to be estimated locally first. A patch-based dissimilarity measure based on this test statistic was used to be more robust in the selection of homogeneous areas \cite{chen2010nonlocal}. The test statistic was also used for weighting the contributions based on an exponential kernel.

Zhong \emph{et al.\ }introduce the nonlocal Lee filter, which is based on combining the shape-adaptive similarity used by the Lee filter with standard patch-based similarities \cite{zhong2013robust}. 
The same likelihood ratio test as in \cite{chen2010nonlocal} is used for setting the NLM weights.
However, this means that both \cite{chen2010nonlocal} and \cite{zhong2013robust} are unable to filter single-look and two-look PolSAR data, as the this test statistic requires that the covariance matrices are not singular.

In \cite{torres2014speckle} a distance between patches is derived under the assumption that the covariance matrix data follow a complex Wishart distribution. Distances are based on divergence measures estimated from patches of covariance matrices, and used to compute weights in nonlocal speckle filtering. A test statistic for equality in distribution of patches is deduced and its p-value derived. The weights are made a piecewise linear function of the p-value, and the filtering can therefore be associated with a user-specified significance level.
Unlike \cite{chen2010nonlocal} and \cite{zhong2013robust}, the stochastic distances nonlocal means (SDNLM) method presented in \cite{torres2014speckle} is able to handle the case where the number of looks $L < 3$.
Contrary to \cite{chen2010nonlocal}, which extends the likelihood ratio test between two covariance matrices to compare two patches, \cite{torres2014speckle} estimates the parameters of the Wishart distribution for the two patches and calculates the divergence between the distributions.

Introduced in \cite{deledalle2014nl}, the NL-SAR algorithm filters SAR images, including polarimetric and/or interferometric data, through a complex framework for nonlocal filtering. The filtering was based on pre-estimates of covariance matrices, where the dissimilarity measure was based on the hypotheses test for equality of covariance matrices under a Wishart distribution assumption from \cite{conradsen2003test}. A specially designed kernel was then used to convert dissimilarities into weights. A bias reduction step was included, where local variance estimates were compared to the expected variance for a homogeneous region, and filtering was reduced for inhomogeneous regions to avoid smearing of bright targets \cite{deledalle2014nl}.
The NL-SAR algorithm is dependant on the manual selection of a "homogeneous area" for learning the kernel. This selection must meet some requirements to be accepted as fully developed speckle by the algorithm and is sensitive to noise correlation in the area.

Hu \emph{et al.\ }test four different similarity measures and kernels for nonlocal filtering of TerraSAR-X data \cite{hu2015non}. Each of these represent one of the four classes of similarity measures categorised in \cite{deledalle2014exploiting}. The proposed kernel used a pair of manually selected homogeneous and heterogeneous areas for estimating the probability density functions (PDFs) for the similarity measure of each patch \cite{hu2015non}. The point where the PDF of the homogeneous area first intersected the PDF of the heterogeneous area was set as a threshold. Covariance matrices whose patch similarity with the position to be estimated was above the threshold were averaged together with a weight of $1$, and others were ignored \cite{hu2015non}.  

Another example of nonlocal filtering of covariance matrices is \cite{xing2017feature}, where a combination of two similarity measures between coherence matrices was used to calculate the weights. One was based on the Pauli decomposition and the other on the coefficient of variation. The two similarity measures were converted to weights by an exponential kernel, and the weights were multiplied when performing the nonlocal averaging of the covariance matrices.

Contrary to \cite{chen2010nonlocal, zhong2013robust, torres2014speckle, deledalle2014nl, hu2015non, xing2017feature}, where the covariance matrices are estimated locally and then averaged nonlocally, we estimate the covariance matrices in a nonlocal way and thus eliminate the need for first calculating them in a pre-estimation step. 

While nonlocal filtering of SAR images has received much attention, guided nonlocal filtering has not. Apart from the works mentioned in Section \ref{subsec:optical-guide-sar-filtering}, there has not been much focus on this approach.
An exception is a framework for guided nonlocal polarimetric covariance matrix filtering that was presented in \cite{ma2018nonlinear}, which used as guide the original noisy polarimetric SAR image itself. The use of the unfiltered input as a guide follows one of the suggestions in the original guided filtering paper \cite{he2012guided}. 
Optical data is generally much less noisy than SAR data and can therefore aid the nonlocal filtering when used as a guide. 

In \cite{giordano2018unmixing} optical data is used to "guide" the decomposition of different scattering mechanisms.
Though this is not strictly filtering or covariance matrix estimation, the application is similar to ours. 
The authors aim to separate (sparse) forest from bare soil by using a coregistered very high resolution ($\SI{0.25}{\metre} \times \SI{0.25}{\metre}$) NDVI image to "unmix" RADARSAT-2 polarimetric matrices corresponding to the two ground cover types. It was found that for mixed land cover, there was no strong correlation between land cover and scattering mechanisms \cite{giordano2018unmixing}. The problem with mixed land cover indicates the challenging nature of our task of canopy state classification for the sparsely forested \fte{}, especially when considering the "damaged" canopy state.

\section{Data collection and preprocessing}
\label{sec:method}

\subsection{The Polmak AOI}
\label{subsec:polmak}

Our main area of interest (AOI) is located near the villages Polmak, Norway, and Nuorgam, Finland (\ang{28}E, \ang{70}N), in an area of the subarctic birch forest which stretches across the Norwegian-Finnish border. The effects on the forest caused by a major geometrid outbreak between 2006 and 2008 are still clearly visible in the form of high stem mortality on both sides of the border \cite{biuw2014long}. 
We have created ground reference datasets for the canopy state on three different scales.  

\begin{figure*}[t] 
  \includegraphics[clip, trim=1.0cm 1.0cm 15.0cm 1.0cm,  width=1.0\linewidth]{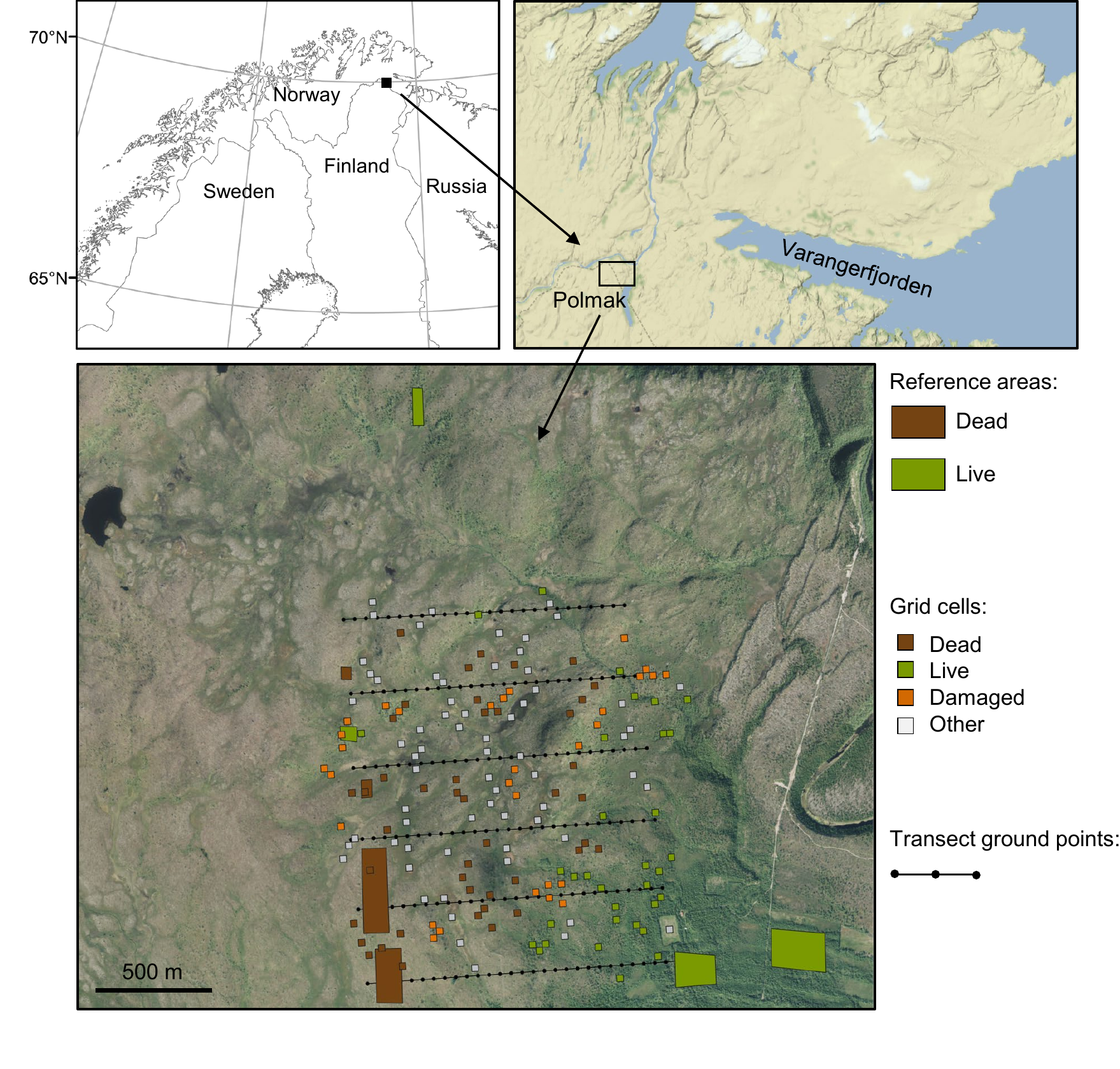}
\caption{The Polmak AOI with the three ground reference datasets. RAs are marked in brown ("dead") and green ("live") rectangles. The classified $\SI{30}{\metre}$ grid cells with "live" cells marked in green, "dead" are dark brown, "damaged" orange, and "na" grey. Finally the six transects are shown as black lines with circles marking the ground plots.}
\label{fig:studysite}
\label{fig:polmak-grid}
\end{figure*}

The AOI was studied during fieldwork on 7-11 August 2017, when 165 ground plots along six transects crossing the Norwegian-Finnish border were examined. The distance between each plot was $\SI{50}{\metre}$, while the distance between each transect was $\SI{200}{\metre}$. The six transects with the 165 ground plots are shown as lines with black dots in Fig.\ \ref{fig:studysite}.
The vegetation was characterised in the $\SI{10}{\metre}\times \SI{10}{\metre}$ ground plots and all trees over $\SI{2}{\metre}$ were measured.
Based on the measured crown area, $\crownarea_\treei$, and proportion of the crown with (live) leaves, $\proplive_\treei \in \left[0, 1 \right]$, for each tree $\treei$, two variables were derived for each ground plot. The total live crown area divided by the ground plot area gave the percentage live canopy, $\text{PLC} = \sum_\treei \frac{\proplive_\treei \crownarea_\treei}{\SI{10}{\metre} \times \SI{10}{\metre}}$. Similarly, the percentage defoliated/dead canopy (PDC) was calculated by replacing $\proplive_\treei$ with $1 - \proplive_\treei$. A simple classification scheme for the canopy state was then employed, where ground plots with $\text{PLC}>5\%$ were defined as "live". Otherwise, if $\text{PDC}>5\%$, the plot was counted as "defoliated". Of the 165 ground plots, 38 were in the "live" class and 57 in the "defoliated" class.

By studying high resolution aerial photographs and comparing images from before (2005) and after the outbreak (2010), eight different-sized reference areas (RAs) were identified, shown as rectangles in Fig.\ \ref{fig:studysite}. 
All RAs were forested before the outbreak, while in four of the eight areas there was no live canopy after the outbreak. These were classified as dead and defoliated forest, marked as brown rectangles in Fig.\ \ref{fig:studysite}. The remaining four RAs, marked in green, constitute the live forest class.
Six of the ground plots examined in 2017 were inside three of the RAs. For these, the classes found from the aerial photographs matched those from the ground plots. Note that the black circles in Fig.\ \ref{fig:studysite} only shows the location of centre of the transect points and do not indicate the extent of the $\SI{10}{\metre} \times \SI{10}{\metre}$ ground plots. 

As an intermediate resolution between the transects with ground plots and homogeneous RAs, a systematic grid of $\SI{30}{\metre}\times \SI{30}{\metre}$ aligned with the transects was laid out over the AOI. Each cell was classified into four classes based on aerial photographs. Three of these were canopy states; "live" (33 cells), "dead" (44 cells), and "damaged" (30 cells). The last class is "other" (66 cells) for vegetation without canopy cover.
We see that the lower left corner of the westernmost live reference area was classified as having damaged canopy state when the grid reference dataset was created.
This is because the coarser labelling of the RAs only consider two classes: "live" and "dead". The inclusion of the intermediate canopy state "damaged" for one of the RAs will potentially have a small negative influence on the classification accuracy for the RA ground reference dataset. However, when extracting the corresponding pixels from the satellite imagery, the number of "damaged" canopy state pixels is small compared to the total size of the RAs. The influence on the classification result is therefore negligible.      

The SAR data consists of two fine-resolution quad-polarisation RADARSAT-2 scenes from 25 July and 1 August 2017 covering the AOI. The scene size is $\SI{25}{\km} \times \SI{25}{\km}$ with a nominal resolution of $\SI{5.2}{\metre} \times \SI{7.6}{\metre}$ (range $\times$ azimuth). 
For the optical guide image, a Sentinel-2 image from 26 July 2017 with no clouds over the AOI was obtained from the Copernicus data hub.

\subsection{Other datasets}
\label{subsec:standard-data}
\label{subsec:flevoland}
\label{subsec:oberpfaffenhofen}

For additional testing of the PGNLM algorithm, two further quad-polarimetric RADARSAT-2 images, formerly provided for testing purposes by MDA, are used. These datasets are from different geographical areas and biomes and contain more varied land cover; making them better suited to illustrate differences in filtering performance, especially when it comes to detail preservation. Landsat-5 images from the same areas were acquired for use as optical guides. Since we required the guide images to be cloud-free, at least over the area corresponding to the SAR scenes, the acquisition dates of the RADARSAT-2 images and the Landsat-5 images are relatively far apart. This is not necessarily problematic for the PGNLM algorithm, since the optical data is only used to help determine the dissimilarities between patches.
In fact, GNLM was shown to be robust to mismatches between the SAR and optical data caused by differences in acquisition times \cite{vitale2019gnlm}. 
Objects appearing or disappearing between the acquisition dates of the PolSAR and optical imagery can cause predictors for the covariance matrix estimate to be de-emphasised or dropped. 
However, as long as the size of the objects is not too large, there will be other predictors available.
In places where large image objects have appeared between the acquisition times, the number and weights of predictors used for estimating the covariance matrices may be low.
For our application of separating live from dead canopy state it is important that the optical imagery is from the summer season when the leaves have sprouted.
This is the case for the data from our AOI described in Section \ref{subsec:polmak}, but for the additional datasets described here this is not important.

The first dataset was acquired over Flevoland in the Netherlands. The RADARSAT-2 imagery was from 2 April 2008.
The Landsat-5 image was acquired 9 September 2008.
The second was from Oberpfaffenhofen, Germany. The acquisition date of the second RADARSAT-2 dataset was 6 April 2008.
For the optical guide, Landsat-5 data from 20 April 2007 was used.

\subsection{Preprocessing}
\label{subsec:preprocessing}

Each SAR dataset was radiometrically calibrated and terrain corrected using the European Space Agency (ESA) Sentinel Application Platform (SNAP) software. For the datasets from Flevoland and Oberpfaffenhofen, the SRTM 1sec HGT digital elevation model (DEM) was used for the terrain correction. This DEM does not cover high latitudes, so for the Polmak AOI the GETASSE30 DEM was used. The output products had $\SI{10.0}{\metre} \times \SI{10.0}{\metre}$ spatial resolution.

For the Sentinel-2 data, atmospheric correction was applied to retrieve top of atmosphere reflectance (TOA). Then the four spectral bands with $\SI{10.0}{\metre}$ spatial resolution, namely red, green, blue, and near infrared (NIR), were extracted. 

For the Landsat-5 data, the six Thematic Mapper (TM) bands with $\SI{30.0}{\metre}$ spatial resolution were selected (TM1, TM2, TM3, TM4, TM5, and TM7). These were resampled to obtain a pixel size of $\SI{10.0}{\metre} \times \SI{10.0}{\metre}$ and the images were coregistered with the SAR data using SNAP.

\section{Results}
\label{sec:results}

\subsection{Setting the parameters}
\label{subsec:param-selection}
\label{sec:pgnlm-params}

The PGNLM parameters for search area and patch sizes were set based on the recommendations for GNLM in \cite{vitale2019gnlm} with necessary modifications. The search area was set to $39 \times 39$ pixels, while the size of the patches to be compared was $5 \times 5$. The latter deviates from \cite{vitale2019gnlm}, which uses a patch size of $8 \times 8$ as a perceived standard in the literature. As mentioned in Section \ref{subsec:nonlocal-covmat-estimation}, we choose a patch size with an odd number of pixels, since for each patch we estimate the covariance matrix for the centre position. 
We also use a smaller patch size due to experiences from the fieldwork done in the AOI, where the fine-scale variations in the landscape do not warrant comparing too large areas. 

The balancing factor between SAR and optical dissimilarities in Eq.\ \eqref{eq:weight-org} was set to $\gamma = 0.85$, while $\lambda$ was $2$. 
As discussed in Section \ref{subsec:normalising-dissims}, this indicates that a predictor whose PolSAR and optical dissimilarities are exactly on the specified percentiles is assigned a weight of $e^{-2} \approx 0.1353$, so $13.5\%$ of the weight which is assigned to the outer product of the target pixel itself.   
For "normalising" the dissimilarities according to Eq.\ \eqref{eq:polsar-tilde-dissim} and \eqref{eq:opt-tilde-dissim}, the 50th percentile (median) was used, i.e.\ $\percentile_\text{Pol} = 50$ and $\percentile_\text{OPT} = 50$.

\begin{figure*}[htb] 
  \centering
  \includegraphics[width=1.0\linewidth, keepaspectratio]{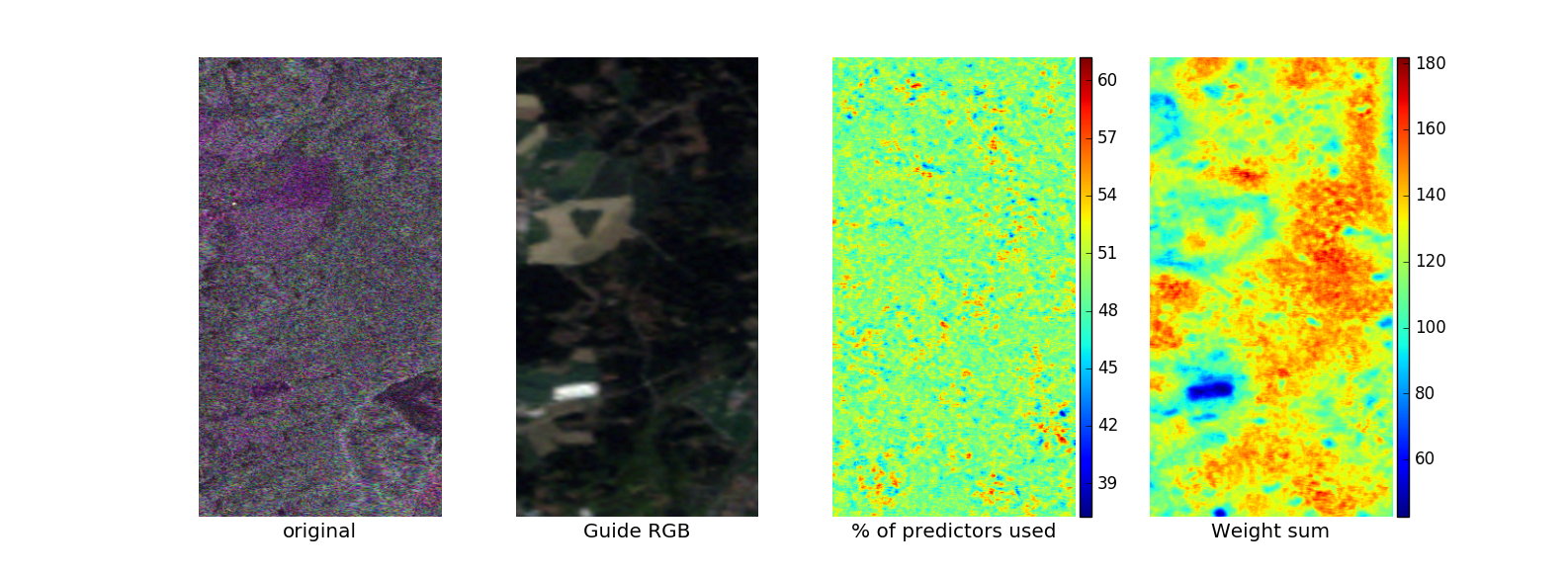}
\caption{Number of predictors and weight sum for an excerpt from the Oberpfaffenhofen dataset. From left to right: Input PolSAR image, corresponding optical guide, percent of predictors used, and sum of weights sum.}
\label{fig:FL4-npatches-used}
\label{fig:npatches-vs-weights}
\end{figure*}
The percentile value for the PolSAR data is also used for discarding unreliable predictors by setting the threshold $T_\text{Pol}$ for reducing the set of all possible predictors $\Omega$ to $\Omega'$ according to Eq.\ \eqref{eq:omega-1prime}. 
The final set of predictors $\Omega''$ is obtained by selecting the $S_0$ predictors with the lowest optical dissimilarity from $\Omega'$, so that the cardinality of the set $\Omega''$ is given by Eq.\ \eqref{eq:omega-2prime-card}. 
To illustrate the effect of the percentile value threshold, we consider the case where we do not limit the maximum number of predictors, that is setting $S_0 = S$. 
The other parameters are as described above, including $\percentile_\text{Pol} = 50$, which means that the set of predictors used will be $\Omega'' = \Omega'$.
Fig.\ \ref{fig:npatches-vs-weights} shows a crop of the Oberpfaffenhofen PolSAR image presented in Section \ref{subsec:oberpfaffenhofen} in the leftmost subplot, with the single-look intensity bands $I_{\text{HH}}$, $I_{\text{HV/VH}}$ and $I_{\text{VV}}$ as the red (R), green (G), and blue (B) channels, respectively. The RGB bands of the optical Landsat-5 guide image are displayed in the middle left subplot. Fig.\ \ref{fig:npatches-vs-weights} then shows
the percentage of predictors used (middle right subplot) and the sum of the weights (right subplot) with the described parameter setting.  

Given the search window size of $39 \times 39$ pixels, the maximum number of predictors available is $S = 1,521$. As expected, since the percentile $\percentile_\text{Pol}$ is set to the median (50th percentile), we see in the third subplot that the percentage of predictors used fluctuates around 50\%. 
This corresponds to the covariance matrices being estimated by the weighted sum of the outer products of $\sim 760$ target vectors.   
The number of predictors used seems independent of the large-scale terrain features of the scene. 
For example, the gravel pit, which appears bright in the optical guide image and can be discerned as a darker rectangular shape in the PolSAR image, does not appear to have a consistently lower (or higher) percentage of predictors used.
Likewise, there is no visible systematic difference for the homogeneous areas of forests and fields.
However, there are small areas where fewer or more predictors are used.
For some pixel locations, up to 60\% ($\sim912$) of the predictors were used, while in the lower end 39\% ($\sim593$) of the available predictors were used.

The plot of the corresponding total weight, as calculated for each pixel position according to Eq.\ \eqref{eq:pgnlm-normconst}, is shown in the rightmost subplot of Fig.\ \ref{fig:npatches-vs-weights}. 
It seems to correspond better with the terrain objects seen in the original speckled PolSAR image and the optical guide than the number of predictors used. Here we can see that the covariance matrix estimates for pixels from forests and fields receive higher sum of weights. This indicates that patches centred on pixels from such image objects have more similar patches in the search area.
We see that the weight sums range from $\sim40$ to $\sim180$ with a mode of around $\sim125$. 
To provide some context for the weight sum, recall that a weight of $w=1$ corresponds to a zero dissimilarity, $\dissim_\text{Pol}= 0$ and $\dissim_\text{OPT}=0$, and unit normalised dissimilarities, $\Tilde{\dissim}_\text{Pol} =\dissim_\text{Pol}/T_\text{Pol}=1$ and $\Tilde{\dissim}_\text{OPT} =\dissim_\text{OPT}/T_\text{OPT}=1$, correspond to a weight of $w=e^{-\lambda}=e^{-2}\approx0.135$.
By combining the information about the number (percent) of predictors with the sum of the weights for the predictors used, we can find the average weight per predictor. 
This varies across the image, as seen from the two subplots, but we can find an exemplar by dividing the mode of the weight sum by the number of predictors when $50\%$ are used. 
Doing this gives a typical average weight per predictor of $\sim0.164$. 
It indicates that in this case, with $760$ predictors and a sum of weights equal to $125$, the majority of the predictors would have a low weight, not much higher than the case where $\Tilde{\dissim}_\text{Pol} = \Tilde{\dissim}_\text{OPT} = 1$ and the weight is $e^{-\lambda} = e^{-2} \approx 0.135$. 
Thus, many of the predictors used would have a relatively high total dissimilarity.
Note that unlike the normalised PolSAR dissimilarity, for which we know that $\Tilde{\dissim}_\text{Pol} \in [0,1]$ due to the thresholding with $T_\text{Pol}$, $\Tilde{\dissim}_\text{OPT}$ does not have an upper bound. 
Any predictor whose patch-wise optical dissimilarity exceeds $T_\text{OPT}$ set by the percentile value $\percentile_\text{OPT} = 50$ would give a dissimilarity value higher than $1$.
One reason for the low average weight in the typical case, could be that a significant number of predictors with high optical dissimilarity (higher than the median) are included in the predictor set $\Omega'$.

\begin{figure*}[htb]
  \centering
  \includegraphics[width=1.0\linewidth, keepaspectratio]{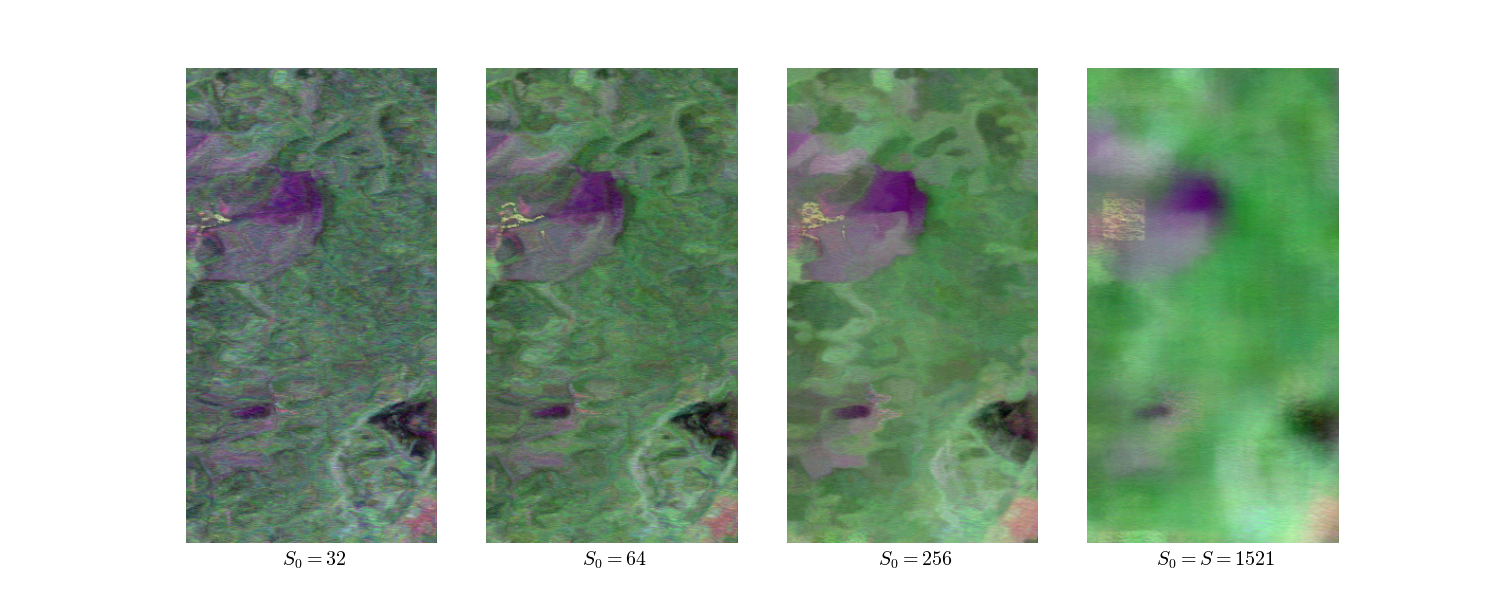}
\caption{Visual quality of the PGNLM estimated covariance matrices for different values of the maximum number of predictors used, increasing from left to right.}
\label{fig:npatches-used-filterres}
\label{fig:s0-comparison}
\end{figure*}

We now consider the second stage of discarding unreliable predictors, where we limit the maximum number to the $S_0$ with the lowest optical dissimilarity. 
The size of the search area, $S$, gives the upper limit of $S_0$.
This parameter must also be set in conjunction with $\percentile_\text{Pol}$, as a higher percentile would increase the threshold $T_\text{Pol}$, which would mean that more predictors satisfy the requirement in Eq.\ \eqref{eq:omega-1prime}. Decreasing the percentile would have the opposite effect. 
So, for the case in Fig.\ \ref{fig:npatches-vs-weights} where the highest number of predictors used was $912$, setting $S_0 > 912$ means that no predictors would be discarded based on the optical dissimilarity as seen from Eq.\ \eqref{eq:omega-2prime-card}.
It should be pointed out that the filtering result corresponding to Fig.\ \ref{fig:npatches-vs-weights}, where $S_0$ has been set equal to $S$, is oversmoothed with significant filtering artefacts and few meaningful details visible.
Fig.\ \ref{fig:npatches-used-filterres} shows images with the diagonal elements of the PGNLM filtered covariance matrices as RGB channels for different values of $S_0$. The diagonal elements, $C_{11}$, $C_{22}$ and $C_{33}$, are the intensity values $I_{\text{HH}}$, $I_{\text{HV/VH}}$, and $I_{\text{VV}}$. 
The rightmost subplot corresponds to the parameter setting used to generate the plots of percent of predictors used and weight sum in Fig.\ \ref{fig:npatches-vs-weights}, with $S_0 = S = 1521$.
We see that the result for $S_0 = 256$ also appears very smooth, and details in forest and fields appear to be missing. The details are also smeared for the village, which appears yellow in the upper left part of the crop.
$S_0 = 64$ appears to give a reasonably smooth result, while preserving details. 
For $S_0 = 32$ there is less smoothing, but this filtered result is much more informative than the original data in the leftmost subplot of Fig.\ \ref{fig:npatches-vs-weights}. 
However, it appears that using a higher number of predictors reveal more details of the village.
The reason could be that the $S_0$ predictors are selected based on the lowest optical dissimilarities $\Tilde{\dissim}_\text{OPT}$, and the optical guide has a spatial resolution $\SI{30.0}{\metre}$, while for the PolSAR data it is $\SI{10.0}{\metre}$, as mentioned in Section \ref{subsec:preprocessing}. 
Although the optical pixels have been resampled to a $\SI{10.0}{\metre} \times \SI{10.0}{\metre}$ grid for coregistration, the resolution does not change. The underlying $\SI{30.0}{\metre} \times \SI{30.0}{\metre}$ resolution may be insufficient to accurately separate between similar patches for the village area.
The PolSAR image has a higher spatial resolution and $\Tilde{\dissim}_\text{Pol}$ may be better suited to weight the dissimilarity of potential predictors, and as the balancing factor $\gamma$ is set to $0.85$, much more emphasis will be put on $\Tilde{\dissim}_\text{Pol}$. 
So in this case, where the maximum number of predictors $S_0$ is low, increasing it will add predictors with a significant weight, even though the optical dissimilarity $\Tilde{\dissim}_\text{OPT}$ for these always will be higher.
The reason is that the predictors selected from $\Omega'$ for $\Omega''$ are not guaranteed to be the ones with the lowest PolSAR dissimilarity, so increasing $S_0$ may actually lead to higher weighted outer products being added to the covariance matrix estimate.
Another factor that potentially speaks against setting the number of predictors too low is a phenomenon described in \cite{vitale2019gnlm} for the GNLM filtering result, where so-called "ghost textures" appear for low values of $S_0$.
It was found that with few predictors the speckle was only partially suppressed, and the averaging of similar patches gave rise to subtle textures that could not be seen in the original data \cite{vitale2019gnlm}.
This problem might not occur in our implementation of the nonlocal filtering, since although the dissimilarity measures are patch-based, the averaging is not.
The parameter settings are also quite different, as \cite{vitale2019gnlm} sets the threshold $T$ such that nearly 100\% of the predictors are used in homogeneous areas.
For our application, where we wish to look at the canopy state in the \fte{}, we set the maximum number of patches used for estimation to $S_0 = 64$, which appears to give a good balance between smoothing and detail preservation.

\subsection{Filtering performance}
\label{subsec:filtering-comparison}
To visually evaluate the filtering performance, we look at two $300 \times 300$ excerpts from the Flevoland dataset. 
With a pixel size of  $\SI{10.0}{\metre} \times \SI{10.0}{\metre}$, these show an area of $\SI{3}{\km} \times \SI{3}{\km}$.  
For comparison with PGNLM we obtain the covariance matrix estimates using the boxcar, refined Lee \cite{refinedlee}, and intensity-driven adaptive-neighbourhood (IDAN) \cite{vasile2006idan} filters implemented in the SNAP software. 
The parameters for PGNLM are described in Section \ref{sec:pgnlm-params}.
Both boxcar and enhanced Lee filters use a $5 \times 5$ window, while the adaptive neighbourhood size for IDAN is set to 50.
We also compare our results with NL-SAR. We download the latest available binaries (v08) for the NL-SAR toolbox from \url{https://www.charles-deledalle.fr/pages/nlsar.php} and use the Python interface. Our data is in the single look complex (SLC) format, so we use the SNAP software to convert the complex target vectors $\mathbf{s}$ to single-look covariance matrices and save in a PolSARpro file format, which is necessary to utilise the NL-SAR toolbox. 
The parameters of NL-SAR are set by considering "a wide variety of parameters and automatically selecting locally the best ones" \cite{deledalle2014nl}.
In practice this works by trying different parameter values and choosing the best ones according to some pre-defined criteria (see \cite{deledalle2014nl} for details).
We use the default values specified in the toolbox documentation, provided at the above mentioned website, for the lower and upper limits that define the "search range" for the automatic tuning of the parameters. 
The limits are specified as follows: the smallest half-width of the search window for nonlocal averaging is $1$ (corresponding to $3 \times 3$ windows) and the largest is $12$ ($25 \times 25$ windows), similarly the half-width of the patches used for calculating the dissimilarity goes from $0$ ($1 \times 1$ "patches") to $5$ ($11 \times 11$ patches).
The "scale" parameter, which defines the half-width of the boxcar averaging window for the local pre-estimate of covariance matrices used for weight computation, is not mentioned in the toolbox documentation. 
We assume that the set of scales consists of $0$ (no averaging), $1$, and $2$ as specified in \cite{deledalle2014nl}.
The algorithm requires manual selection of a homogeneous area from which to calculate the noise statistics. We select areas of $100 \times 100$ pixels from each of the scenes.
The kernel for mapping dissimilarities to weights for nonlocal averaging is derived from this homogeneous region. The mapping includes an exponential function with a kernel width parameter $h$, but \cite{deledalle2014nl} argues that it can be be kept constant for different inputs as the mapping is learned from the data.
We use the default value in the toolbox implementation: $h = 1.0$.

\begin{figure}[htb]
\begin{minipage}[b]{1.0\linewidth}
  \centering
  \includegraphics[width=1.0\linewidth, keepaspectratio]{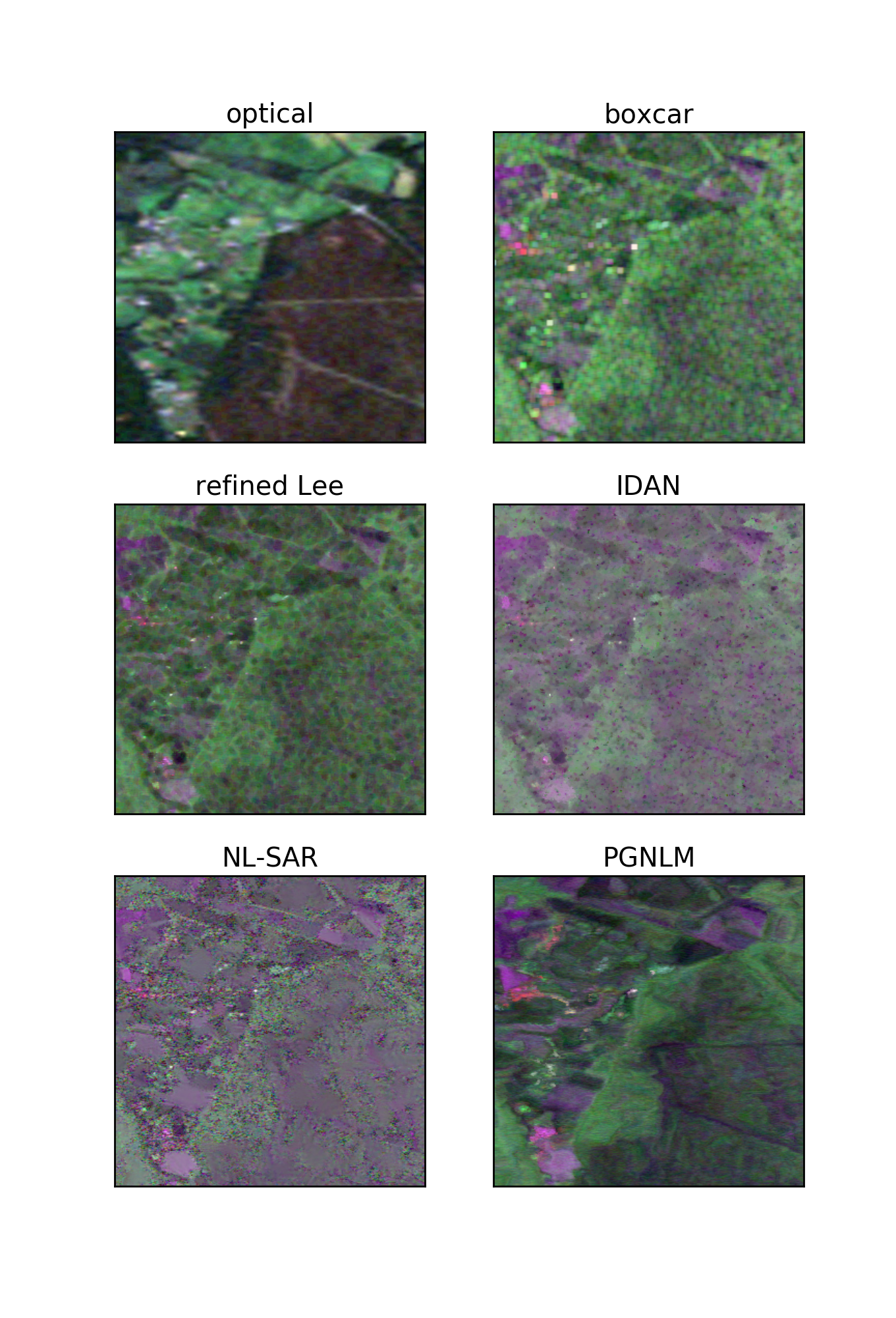}
\end{minipage}
\caption{First example from Flevoland scene: Optical Landsat data (top left corner) and PolSAR data processed with a boxcar filter (top right), refined Lee filter (left middle), IDAN filter (right middle), NL-SAR filter (left bottom), and the proposed PGNLM algorithm (right bottom).}
\label{fig:FL5-res0-compare-6}
\label{fig:compare-6-res1}
\end{figure}

The first example in Fig.\ \ref{fig:compare-6-res1} shows a small segment of the Flevoland dataset. It is chosen so that it contains a mix of land-cover types; buildings, roads, fields, bare ground, and forest. For the PolSAR data, $C_{11}$, $C_{22}$, $C_{33}$ are used as the red, green, and blue channels, respectively.
Each channel of the excerpts (including the optical RGB image) is normalised separately to have values between $0$ and $1$.
We see that that PGNLM algorithm manages to clearly preserve both roads that cross the bare soil, visible in the lower right part of the optical image, whereas the other filtering methods struggle with the top road. PGNLM also appears to preserve other details while having a smooth appearance and good contrast between the different ground covers.

\begin{figure}[htb]
  \centering
  \includegraphics[width=1.0\linewidth, keepaspectratio]{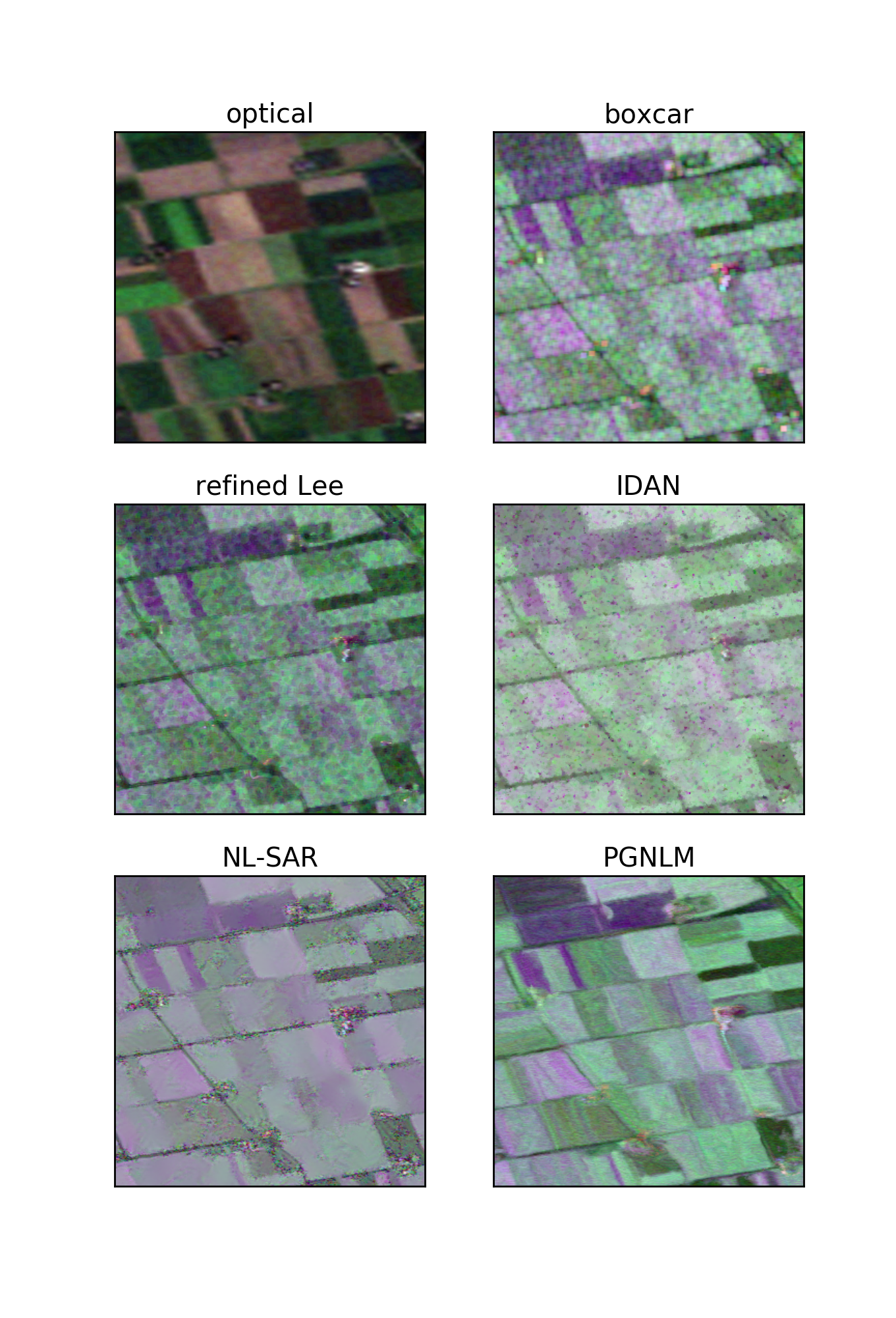}
\caption{Second example from Flevoland scene: Optical Landsat data (top left corner) and PolSAR data processed with a boxcar filter (top right), refined Lee filter (left middle), IDAN filter (right middle), NL-SAR filter (left bottom), and the proposed PGNLM algorithm (right bottom).}
\label{fig:FL5-res2-compare-6}
\label{fig:compare-6-res2}
\end{figure}

Fig.\ \ref{fig:FL5-res2-compare-6} covers another part of the Flevoland dataset, focusing on different types of agricultural fields. 
The PGNLM algorithm achieves a significantly smoother result than the other filtering methods, and without any obvious filtering artefacts. This is not unexpected, as it averages covariance matrix estimations from a large search area.
We note that the road that goes from the top left part of the excerpt towards the lower middle is not as sharply contrasted in our filtering result as in the NL-SAR or refined Lee results.
A possible explanation for this is that, unlike the roads in Fig.\ \ref{fig:FL5-res0-compare-6}, this road is not visible in the $\SI{30.0}{\metre} \times \SI{30.0}{\metre}$ resolution Landsat-5 image.
Hence, the optical guide will not be helpful in filtering the road pixels.
The guide is, however, very helpful in differentiating between the different types of fields, as can be seen from the PGNLM filtering result.
The different colour nuances can be clearly seen in the upper left subplot of Fig.\ \ref{fig:FL5-res2-compare-6}. Furthermore, in addition to the RGB bands displayed, the guide image contains three additional bands (TM4, TM5, and TM7) that are useful for differentiating between the various crop types.

\subsection{Classification results for AOI}
\label{subsec:AOI-results}
We will now look at canopy state classification for our area of interest using PolSAR data and assess how our covariance matrix estimation method influences the result compared to standard filtering methods. 
We extract pixels from the satellite imagery for the three different ground reference datasets described in Section \ref{subsec:polmak}. 
Based on these we train a random forest classifier with 200 trees on the filtered covariance matrices. The random forest algorithm is implemented in the Scikit-learn Python package \cite{scikit-learn}. 
All the polarimetric information is contained in the elements of $\mathbf{C}$, and we can further simplify the processing by extracting the available features that will be used in the classification: $C_{11}$, $C_{22}$, $C_{33}$, $|C_{13}|$, and $\angle C_{13}$. Here, $C_{11}$, $C_{22}$, $C_{33}$ are the intensities in the HH, HV, and VV channels, respectively, and $C_{13}=|C_{13}|e^{j\angle C_{13}}$ is the cross-correlation between the complex scattering coefficients in the co-polarised channels HH and VV. Covariance matrix elements $C_{12}$ and $C_{23}$ may contain information about the slope of the terrain, but should theoretically not contain any information useful for the discrimination of different vegetation types or properties, and are therefore omitted. We compare the classification results obtained with different algorithms for covariance estimation and filtering.
In addition, we compare with the classification result obtained on a feature vector containing the four Sentinel-2 (S2) bands used as the guide in PGNLM. 
We divide the data into four non-overlapping parts for $k$-fold cross validation using Scikit-learn \cite{scikit-learn}, where one part (fold) is used for testing and the remaining folds are used for training. This is repeated, so each of the $k = \crossvalk$ folds are used for testing. 

\begin{figure}[htb]
  \centering
  \includegraphics[width=1.0\linewidth, keepaspectratio]{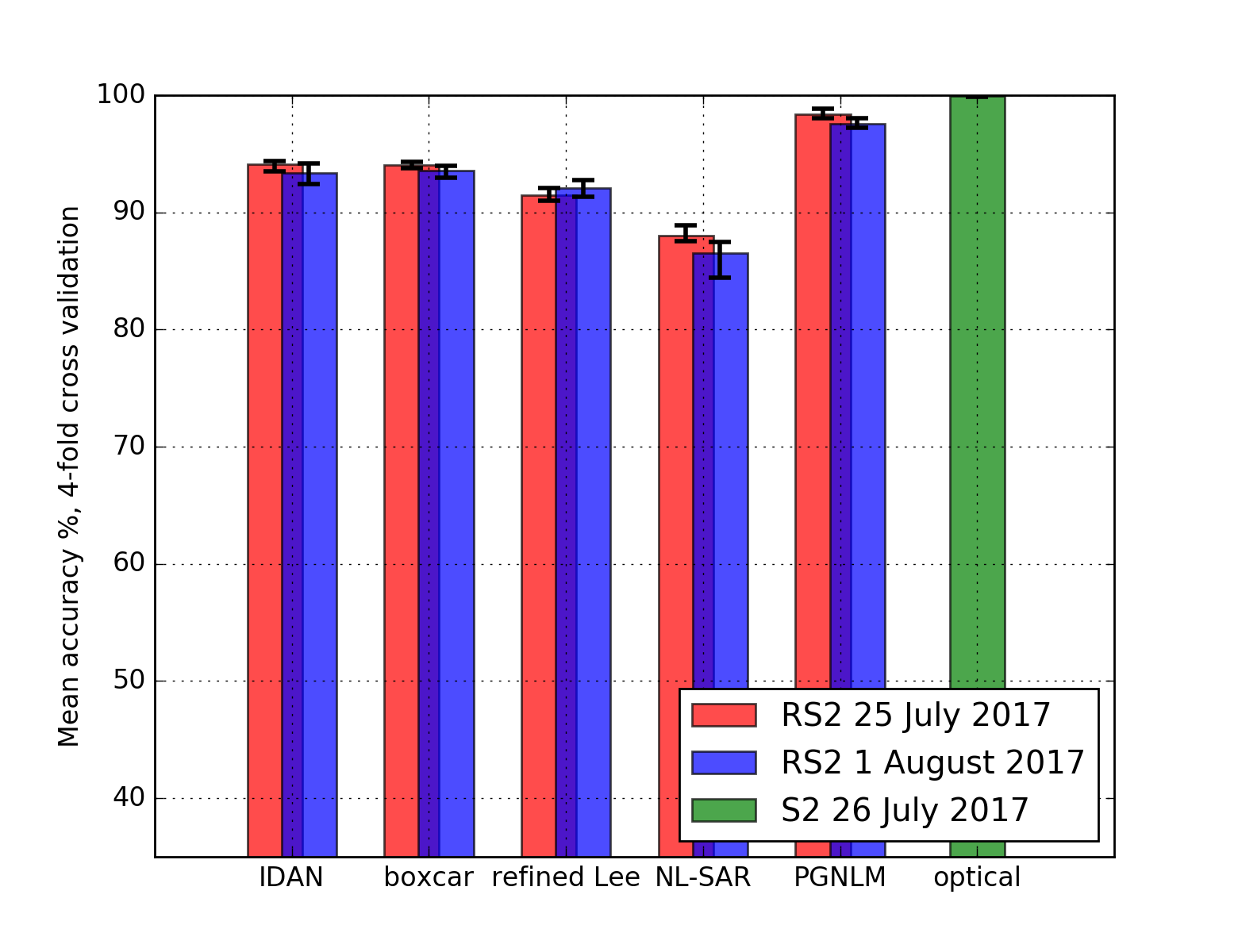}
\caption{Classification accuracy for reference areas (RAs).}
\label{fig:ra-accuracy-bars}
\end{figure}

We first consider the classification on pixels extracted from the eight homogeneous reference areas (RAs) for the "live" and "dead" canopy states seen in Fig.\ \ref{fig:studysite}. 
Fig.\ \ref{fig:ra-accuracy-bars} shows the average classification accuracy across the $k$ folds plotted as solid bars. The minimum and maximum accuracy for the four folds are show as black lines for each average accuracy. 
Fig.\ \ref{fig:ra-accuracy-bars} and the other bar plots in this section have the same y-axis to enable direct comparison between the results for the three ground reference datasets. The baselines have been set to $35\%$ to ease visual comparison of the different classification accuracies.
We see that PGNLM achieves 98.4\% accuracy for 25 July (red bars), 4.3 percentage points better than the second best (IDAN). For the 1 August dataset (blue bars) PGNLM achieves 97.5\% accuracy, 4.0 percentage points better than the  second best (boxcar). The optical data (Sentinel-2 26 July) has an 99.9\% classification accuracy shown as indicated by the green bar.
This is a relatively simple classification task for several reasons. 
Firstly, the dataset is large with a total of 1720 pixels from the eight RAs.
Secondly, the two canopy states are clearly visible from the above using the high resolution aerial photographs. 
A third factor which should be mentioned is that each of the four folds will most likely contain pixels from all eight RAs.
This also explains why there is very little variation in minimum and maximum accuracy for the four folds.

\begin{figure}[htb]
  \centering
  \includegraphics[width=1.0\linewidth, keepaspectratio]{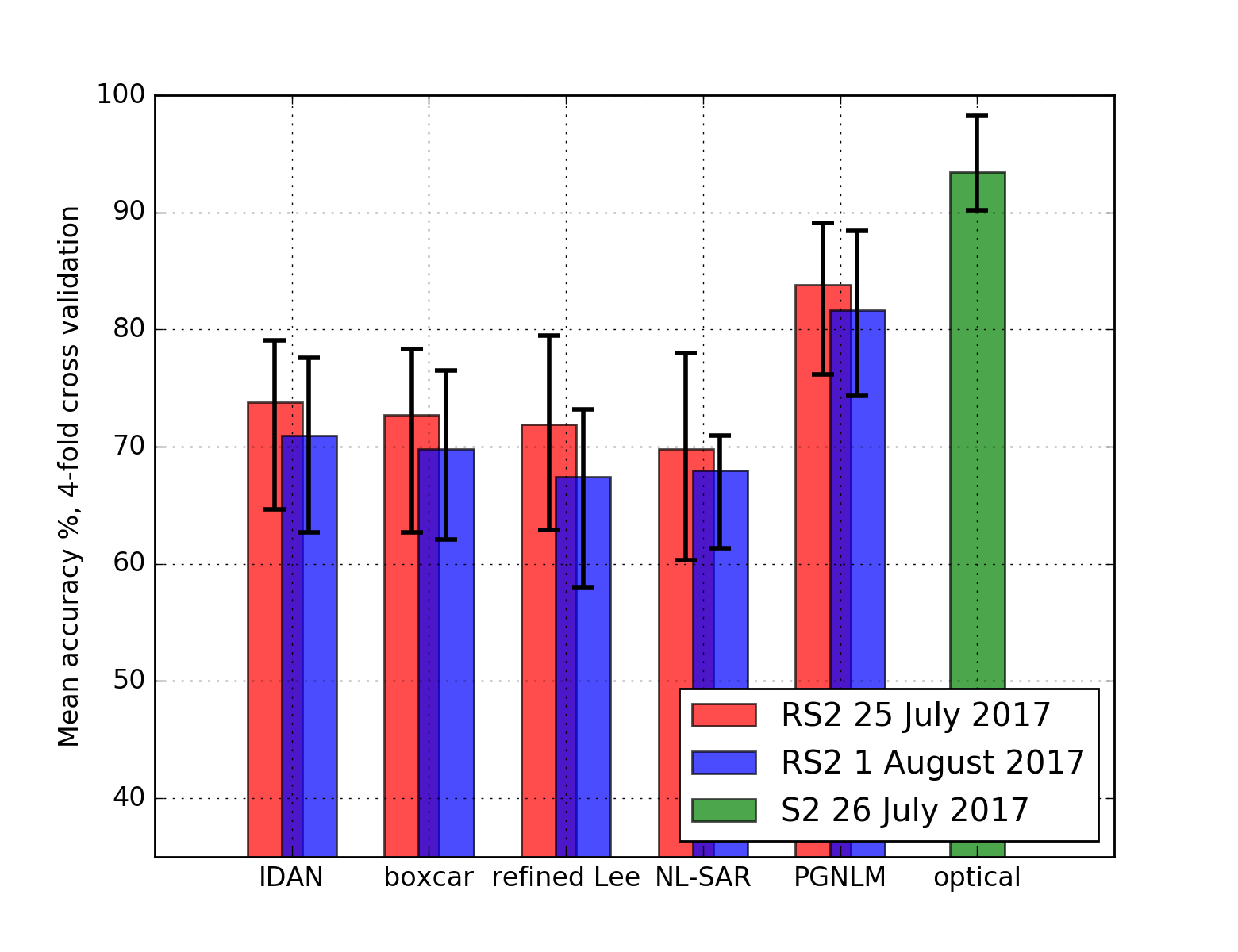}
\caption{Classification accuracy for grid cells.}
\label{fig:grid-accuracy-bars}
\end{figure}

The grid shown in Fig.\ \ref{fig:polmak-grid} can be seen as an intermediate spatial scale between the transect ground plots and the homogeneous reference areas. 
To avoid training and testing on pixels from the same grid cell, we use group $k$-fold cross-validation from Scikit-learn \cite{scikit-learn} with four folds. 
The pixels from each grid cell will then be grouped together in one fold and will therefore not be in both the training and the test set of the cross-validation. 
The parameters of the RF classifier are the same as before.
We see that the classification accuracies in Fig.\ \ref{fig:grid-accuracy-bars} are lower than for the RAs in Fig.\ \ref{fig:ra-accuracy-bars}, as expected from the perceived difficulty of this classification task.
The Sentinel-2 data achieves 93.5\% average accuracy. The PGNLM-estimated covariance matrices again achieve the highest classification accuracy for the PolSAR data with 83.9\% and 81.7\% for the 25 July and 1 August datasets, respectively. The IDAN-filtered covariance matrices have the second best result with 73.8\% and 71.0\%, more than 10 percentage points lower than PGNLM.

The labelled grid cells also contain two other classes of interest, namely "damaged" canopy state and "other" vegetation. In Table \ref{tab:grid-other-classes} we summarise the average  classification accuracies for the different covariance matrix estimates and the four-band Sentinel-2 dataset when these two classes are included. 
The classifier and group $k$-fold cross-validation use the same setup as for the classification of "live" versus "dead" canopy state.
We also run the classification on the combination of the five PolSAR features from the PGNLM-estimated covariance matrices with the four Sentinel-2 bands that were used as the guide image.
\begin{table*}[t]
    \begin{center}
        \begin{tabularx}{0.8\linewidth}{ c c c c c c c } 
            \hline
            Filtering & 25/07 + damaged & 01/08 + damaged  & 25/07 + other & 01/08 + other  & 25/07 + both & 01/08 + both  \\
            \hline
            IDAN         &  51.1 & 54.4  & 54.9 & 52.2  & 46.0 & 42.4  \\ 
            boxcar       &  50.7 & 52.7  & 55.8 & 53.4  & 46.2 & 41.8  \\ 
            refined Lee  &  50.0 & 48.8  & 55.6 & 49.7  & 45.4 & 38.4  \\
            NL-SAR       &  47.2 & 45.2  & 50.0 & 50.5  & 40.8 & 40.1  \\
            PGNLM        &  58.4 & 60.3  & 61.1 & 58.7  & 51.0 & 48.3  \\
            \hline
            optical (26/07) &  68.6 & 68.6  & 73.5 & 73.5  & 59.5 & 59.5  \\
            combination  &  73.5 & 70.7  & 79.2 & 78.2  & 66.9 & 64.0  \\
            \hline
        \end{tabularx}
    \end{center}
    \caption{Classification accuracy (\%) on grid cells for the 25 July (25/07) and 1 August (01/08) datasets, when in addition to the "live" and "dead" classes, the "damaged" class, "other" class, or both are included.}
    \label{tab:grid-other-classes}
\end{table*}
We see from Table \ref{tab:grid-other-classes} that the accuracy drops significantly for all datasets for all cases, compared to the simpler "live" versus "dead" canopy state classification baseline shown in Fig.\ \ref{fig:grid-accuracy-bars}.
The result on the optical data is again the best, regardless of which of the two RADARSAT-2 datasets it is compared to.
Also in this case the classification accuracies for the PGNLM-estimated covariance matrices are higher (between 4.8 and 7.3 percentage points) than the second best filtered PolSAR result. 
The combination of the PolSAR and optical features gives a noticeable increase (from 2.1 to 7.4 percentage points) in the average classification accuracies for all cases.
The accuracy also increases in the case where optical alone achieves a quite high classification accuracy.
For the two-class setting in Fig.\ \ref{fig:grid-accuracy-bars} the average accuracy increases from 93.5\% to 95.2\% and 95.0\% when adding the PGNLM 25 July and 1 August data respectively.
Thus it is clear that the PolSAR data contains features relevant for all classes that are not captured by the optical data. 

\begin{figure}[htb]
  \centering
  \includegraphics[width=1.0\linewidth, keepaspectratio]{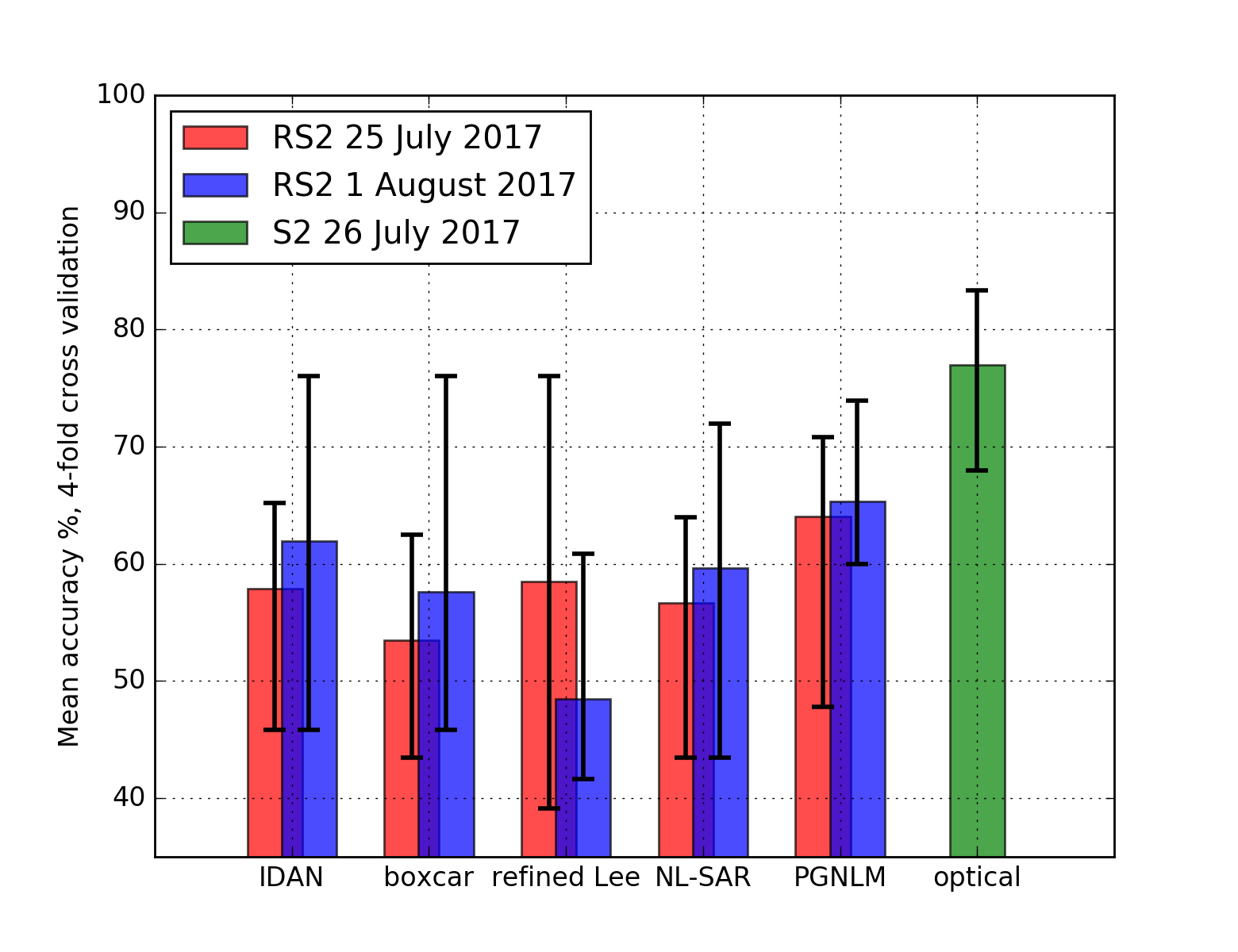}
\caption{Classification accuracy for transect ground plots.}
\label{fig:accbars}
\label{fig:transect-accuracy-bars}
\end{figure}

The classification result for the $\SI{10.0}{\metre} \times \SI{10.0}{\metre}$ transect ground plots described in Section \ref{subsec:polmak} is shown in Fig.\ \ref{fig:transect-accuracy-bars}.
The RF parameters and cross-validation setup are the same as for the result in Fig.\ \ref{fig:grid-accuracy-bars}.
We see that all accuracies are lower than for the grid cells in Fig.\ \ref{fig:grid-accuracy-bars} and significantly lower than for the RAs in Fig.\ \ref{fig:ra-accuracy-bars}.
The average classification accuracy for the optical data is 77.0\%, significantly higher than for any results based on the filtered PolSAR covariance matrices. Of these, the PGNLM achieves the highest average accuracy with 64.1\% for the 25 July dataset and 65.4\% for 1 August. The second highest PolSAR accuracies are 58.5\% for 25 July (refined Lee) and 62.0\% for 1 August (IDAN).
We see from the black bars in Fig. \ref{fig:transect-accuracy-bars} that the minimum and maximum classification accuracies have large variations for most of the datasets.
Part of the explanation for this is that the dataset is quite small, as described in Section \ref{subsec:polmak}, with 38 samples from the "live" class and 57 from the "defoliated/dead" class, each corresponding to a single pixel extracted from the satellite imagery.
It should be noted that this is a challenging classification task, particularly because co-registration of the transect ground plots and the satellite dataset is difficult. Fine-quad RADARSAT-2 data has an absolute error in geolocation of $<\SI{15.0}{\metre}$ \cite{giordano2018unmixing}, and while the centre of each ground plot was recorded with a differential global positioning system (GPS) instrument, the measurement of the $\SI{10.0}{\metre} \times \SI{10.0}{\metre}$ extent was not as accurate. Thus, it is not a perfect overlap between the pixels and ground plots. This can be problematic, since a single tree anywhere within the ground plot can be sufficient to label it as either "live" or "defoliated/dead". 12 out of a total 95 ground plots used were labelled based on a single tree according to the classification scheme used. 
The grid cells, on the other hand, are labelled based on canopy state changes that are visible from aerial photographs, and cover $\SI{30.0}{\metre} \times \SI{30.0}{\metre}$.
Thus, at least some of the $\SI{10.0}{\metre} \times \SI{10.0}{\metre}$ pixels will fully overlap with the grid cell.
We also avoid potential issues that may be caused by the class label being defined by a single tree.

Somewhat surprisingly we see that the other nonlocal method, NL-SAR, is consistently amongst the worst of the reference methods for the canopy state classification. 
Visual inspection of the excerpt from our AOI shows that the NL-SAR result appears to be the least smooth of all the comparison methods.
We can see similar tendencies of relatively poor speckle suppression in the forested areas along the edge of the "V"-shape in the lower left part of Fig.\ \ref{fig:FL5-res0-compare-6}. 
However, the man-made structures appear well preserved, and some fields appear to be well smoothed, which is also the case for the fields seen in Fig.\ \ref{fig:FL5-res2-compare-6}.
We hesitate to make any conjectures about what might cause the apparent lack of smoothing for forested areas seen in NL-SAR results for our data, but note that the PGNLM algorithm has the advantage of using an optical guide to help select the best predictors. Furthermore, since the dissimilarity measure of PGNLM is not based on covariance matrices, we do not need to make provisions for handling the case when the empirical covariance matrix is singular in the case of single-look input, as was done in \cite{deledalle2014nl}. Finally, while both PGNLM and NL-SAR decrease the weights used for nonlocal averaging with increasing dissimilarity, PGNLM benefited greatly from combing this with limiting the number of predictors used as shown in Fig.\ \ref{fig:npatches-used-filterres}.

\subsection{Effect of the optical guide}
\label{subsec:ablation}
In Section \ref{subsec:param-selection} we set the balancing factor $\gamma = 0.85$, which means that in Eq.\ \eqref{eq:weight-pgnlm}, $\Tilde{\dissim}_\text{Pol}$ is weighted by $0.85$ while $\Tilde{\dissim}_\text{OPT}$ is weighted by $(1 - \gamma) = 0.15$. As discussed in Section \ref{subsec:normalising-dissims}, these dissimilarities are roughly on the same scale, so this indicates that the PolSAR dissimilarity contributes significantly more to the weights. Therefore, it is worth investigating how much influence the optical guide has on the covariance matrix estimates. 
In addition to its contribution to the weights, the optical dissimilarity plays an important role in discarding unreliable predictors when limiting the maximum number of predictors to $S_0 \leq S$.
The subset of predictors used, $\Omega''$, is selected from those that fulfil the criterion in Eq.\ \eqref{eq:omega-1prime}, as those with the lowest patch-wise optical dissimilarity $\Tilde{\dissim}_\text{OPT}$.
Thus, if we set $S_0 = S$ (using all predictors), the guide will only contribute through the weights. 
Then, by setting $\gamma = 1.0$ in Eq.\ \eqref{eq:weight-pgnlm}, we can obtain a nonlocal estimate of the covariance matrices without any influence of the optical guide. Eq.\ \eqref{eq:weight-pgnlm} is then reduced to $w(\filti,\centi) = \exp \left[  - \lambda \Tilde{\dissim}_\text{Pol}(\filti,\centi) \right]$.

The importance of discarding unreliable predictors by limiting the maximum number used was illustrated in Fig.\ \ref{fig:npatches-used-filterres}. Therefore, a fair comparison of the results obtained with and without the optical guide requires that we also limit the number of predictors in the latter case.
A natural choice is then to use the $S_0$ predictors with the lowest PolSAR dissimilarity $\Tilde{\dissim}_\text{Pol}$.
These dissimilarities also need to be below the threshold $T_\text{Pol}$. 
In the Oberpfaffenhofen example shown in Fig.\ \ref{fig:npatches-vs-weights}, the smallest number of predictor patches with dissimilarity below the threshold was significantly higher ($\sim600$) than the maximum number used, $S_0 = 64$.

\begin{figure}[htb]
  \centering
    \includegraphics[width=1.0\linewidth, keepaspectratio]{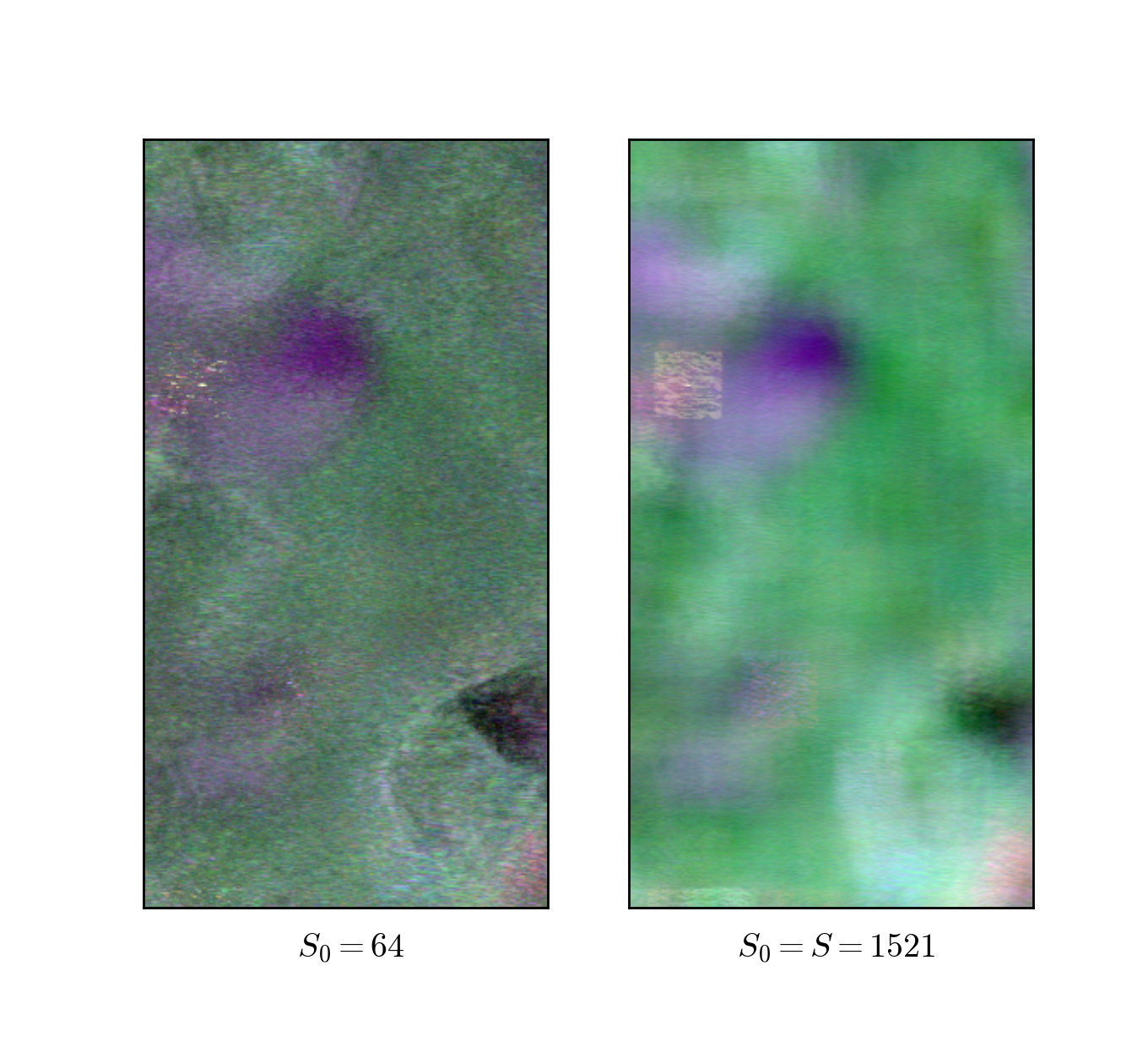}
\caption{Visual quality of covariance matrix estimates in an excerpt of the Oberpfaffenhofen scene obtained without optical guide and respectively with (left) and without (right) a limit on the maximum number of predictors.}
\label{fig:ablation}
\label{fig:no-opt}
\end{figure}

Fig.\ \ref{fig:no-opt} shows the result of the covariance matrix estimate for the two cases discussed above. The result with a maximum of $S_0 = 64$ predictors, selected as the ones with lowest $\Tilde{\dissim}_\text{Pol}$, is shown in the left subplot of Fig.\ \ref{fig:no-opt}. The right subplot shows the result when all predictors in $\Omega'$ (where $\Tilde{\dissim}_\text{Pol} \leq 1.0$) are used. In both cases, the weights are solely based on the PolSAR dissimilarity, and apart from $S_0$ and $\gamma = 1.0$, the parameters are the same as in Section \ref{subsec:param-selection}. 
To assess the impact of the optical guide, Fig.\ \ref{fig:no-opt} should be compared with the corresponding results in Fig.\ \ref{fig:s0-comparison}. 
In particular, it is enlightening to compare the result when $S_0 = S = 1521$ with the corresponding result in the rightmost subplot of Fig.\ \ref{fig:npatches-used-filterres}, as there is no discarding of predictors according to the maximum number set in Eq.\ \eqref{eq:omega-2prime-card}.
Thus, the comparison of $S_0 = S$ between Fig.\ \ref{fig:no-opt} and Fig.\ \ref{fig:s0-comparison} is not influenced by the difference in how the $S_0$ predictors are selected, and solely shows the effect of increasing $\gamma$ from $0.85$ to $1.0$, when the optical dissimilarities do not influence the weights.
Although at first glance the two results seem quite similar, both being significantly oversmoothed,
there is a noticeable difference when we study the lower left part of the image.
The gravel pit, which is visible as the dark area in the lower part of the original PolSAR image and the bright area in corresponding location of the optical guide image of Fig.\ \ref{fig:npatches-vs-weights}, is not discernible for $S_0 = S$ in Fig.\ \ref{fig:no-opt}.
This is in contrast to the result for $S_0 = S$ in Fig.\ \ref{fig:s0-comparison}, where this image object has been preserved.
The bright, homogeneous appearance of the gravel pit in the optical guide means that $\Tilde{\dissim}_\text{OPT}$ will be low between patches that lie within this object and high between patches on the inside and on the outside of the pit. 
Thus, the centre pixels of the former will be weighted more than those from the latter when the outer products are averaged to form the covariance matrix.
As seen from this comparison, the guide in this case helps preserve an object which is visible in both the optical imagery and the single-look intensity PolSAR image shown in Fig.\ \ref{fig:npatches-vs-weights}.

We see by comparing the subplots of Fig.\ \ref{fig:ablation} that the result when limiting the number of predictors to the $S_0 = 64$ with the lowest PolSAR dissimilarity preserves more detail than when the maximum number of predictors are not limited.
However, it appears considerably more blurry than the corresponding result for $S_0 = 64$ in Fig.\ \ref{fig:s0-comparison}.
Again, the gravel pit highlights the difference between the results with and without the optical guide.
While this object is well preserved when the guide is used in Fig.\ \ref{fig:s0-comparison}, it is barely discernible for $S_0 = 64$ in Fig.\ \ref{fig:ablation}.
Clearly this object is difficult to preserve based on the PolSAR dissimilarities alone, and the optical image is fulfilling its role as a guide for selecting and weighting the predictors. 
For this example, the contribution from the optical guide is important, both for setting the weights in the covariance matrix estimate, and for selecting the best predictors.

We now look at how the absence of the optical guide affects the canopy state classification accuracy. 
We use the same RF classification and cross-validation setup as in Section \ref{subsec:AOI-results} for the easiest (RAs) and most challenging (all four classes of the grid cells) classification tasks. 
Based on the visual analysis in Fig.\ \ref{fig:ablation}, we opted to compare the results for nonlocal covariance estimation with the same number of predictors ($S_0 = 64$) as used for the standard PGNLM results in Section \ref{subsec:AOI-results}, as these seem most comparable.
In addition we have included the classification accuracy for the original speckled PolSAR single-look intensity data. Recall that the classification of covariance matrices is based on the five features $C_{11}$, $C_{22}$, $C_{33}$, $|C_{13}|$, and $\angle C_{13}$. For the single-look intensity data, only the intensities are available.
The classification results for the original data and covariance matrices estimated without the guide are summarised in Table \ref{tab:ablation}. For comparison the corresponding PGNLM results are repeated.
\begin{table}[h]
    \begin{center}
        \begin{tabularx}{\linewidth}{ c c c c c  } 
            \hline
            Filtering & 25/07 RAs & 01/08 RAs & 25/07 grid & 01/08 grid  \\
            \hline
            no filtering           &  68.0 & 65.7  & 34.3 & 34.7  \\ 
            no guide            &  97.3 & 96.0  & 47.1 & 44.8  \\ 
            PGNLM               &  98.4 & 97.5  & 51.0 & 48.3  \\
            \hline
        \end{tabularx}
    \end{center}
    \caption{Classification accuracy (\%) on the RAs and grid cells for the 25 July (25/07) and 1 August (01/08) datasets.}
    \label{tab:ablation}
\end{table}

The classification accuracy for the original data is significantly lower than the unguided and standard PGNLM result. 
It is also significantly lower than the other covariance matrix filtering results, as can be seen by comparing Table \ref{tab:ablation} with Fig.\ \ref{fig:ra-accuracy-bars} for the reference areas and Table \ref{tab:grid-other-classes} for the grid cells. 
This is as expected, both because of the lack of speckle suppression, and since the classification based on the original data uses fewer features.
The results for the unguided nonlocal covariance matrix estimates are some percentage points lower than for the corresponding PGNLM results.
However, the accuracies are in general quite high, and better than the results for the other filtering methods.

All in all, for our classification problem, the use of features from the polarimetric covariance matrices is clearly better than the use of the intensity data directly, and the nonlocal estimation of the covariance matrices is better than the other methods. 
The optical guide contributes not only to the perceived visual quality of the filtering result, but also to the accuracy of our classification of canopy state.

\section{Conclusions and future work}
\label{sec:conclusion}
The PGNLM covariance matrix estimation algorithm achieves a good balance between preservation of polarimetric information and spatial resolution on the one hand, and speckle suppression on the other.
Several factors contribute to this: the use of patch-based dissimilarities and nonlocal averaging has proven successful in previous work, and is once again a success factor. 
The optical guide image is important, as it not only contributes to a better setting of weights, but also to discarding unreliable predictors.
In conjunction with the thresholding based on the PolSAR dissimilarity, we ensure that only the most similar predictors are used.
However, contrary to previous works we do not filter pre-estimated covariance matrices, but instead estimate them directly from the SLC-formatted PolSAR product.  

For our application of separating between different canopy states in the \fte{}, we see a significant improvement in the classification accuracy compared to standard PolSAR filtering methods.
Considering the large, homogeneous reference areas (RAs), the classification accuracy on the features from the covariance matrices estimated with our method is almost on par with that of the $\SI{10.0}{\metre}$ resolution Sentinel-2 bands.
Although this is a relatively simple classification problem, our method still achieves an improvement of 4 percentage points or more compared to the other SAR results. 
When we look at classifying the grid cell data we see that the difference between the result for optical data and PGNLM covariance matrices increases, as does the difference between PGNLM and the second best SAR results.
One should be careful with drawing conclusions based on the results for the transect ground plots, but we see that the general trend of optical yielding the highest accuracy followed by PGNLM continues.
Even so, the ability to classify the canopy state of a single pixel represents an aspirational goal that, if successful, would provide ecologists studying the \fte{} with valuable information.
In situ measurements represents the "gold standard" of ground reference data, and ground plot characterisation such described in Section \ref{subsec:polmak} could provide the opportunity to investigate the potential for assessing the degree of canopy damage as a continuous variable.
However, this requires the field work protocol to be designed in such a way that the ground reference data collected is robust to the misalignment with the satellite imagery and geolocation inaccuracies.

We also see that combining the five polarimetric features extracted from the PGNLM covariance matrices with the four optical bands gives a noticeable improvement in the accuracies for all the classification task subsets.
While the combination of data from different modalities, such as SAR and optical, is a tempting approach to improve the results, this requires the datasets to be relatively close in time.
The phenology of the study site, which is itself a promising information source to consider for improving the classification, must be taken into account when the goal is to classify the canopy state.
However, obtaining cloud-free and affordable optical data still proves to be problematic, even after the launch of several new satellites, which prompted this study of using polarimetric SAR data.
PGNLM does not require the SAR and optical guide to be close in time, which means we can take advantage of whatever optical data that can be obtained.

The PGNLM algorithm contains quite a few parameters, that in various ways impact each other. Here they were set in a heuristic manner and kept constant for all experiments. 
However, we believe we have simplified the parameter selection significantly by setting the threshold for discarding predictors based on the SAR domain dissimilarity as a percentile $\percentile_\text{Pol}$, rather than the actual threshold value $T_\text{Pol}$.
Likewise, by dividing the optical dissimilarity values by a percentile value, we have more control over its range, and the balancing factor $\gamma$ does not need to be adjusted for the sensor-dependent or scene-dependent scale of the dissimilarity measures.
The kernel scale parameter $\lambda$ should also be easier to set when we see it in connection with the percentiles we have set, as suggested in this paper. 
We can consider a hypothetical case of a predictor whose PolSAR and optical dissimilarities are exactly at the threshold values, and how much weight should be assigned to that predictor compared to the "centre predictor" (the outer product of the target vector at the pixel position to be estimated).
For future work, it could be considered to drop the balancing factor $\gamma$, as the $\percentile_\text{OPT}$ only regulates the optical dissimilarities $\Tilde{\dissim}_\text{OPT}$ in the weight function, and thus implicitly the balancing with the PolSAR dissimilarity.
Regardless, the combination of parameters of the algorithm should be studied to better understand how they interact.
Synthetic datasets could be used for this purpose as they represent a controlled setting.
Also, applying PGNLM to datasets from other SAR sensors can help get a more accurate comparison of its performance relative to other polarimetric speckle filtering methods.
This can also test our hypothesis that normalising the dissimilarities based on percentile values makes parameter configuration easier, as ideally the same percentiles should work for PolSAR data from other sensors.

We have eschewed traditional measures of measuring filtering performance and instead focused on analysis of the classification results in our target application, which is what motivated us to pursue the nonlocal covariance matrix estimation in the first place. For numerical comparison with different filtering methods, measures such as the equivalent number of looks (ENL) could be calculated. 

Currently the algorithm is implemented in Python for testing purposes, but in order to make it operational, the code should be optimised to speed up processing.

\bibliographystyle{IEEEtran}
{\small
\bibliography{phdref.bib}}

\vskip -2\baselineskip plus -1fil
\begin{IEEEbiography}[{\includegraphics[width=1in,height=1.25in,clip,keepaspectratio]{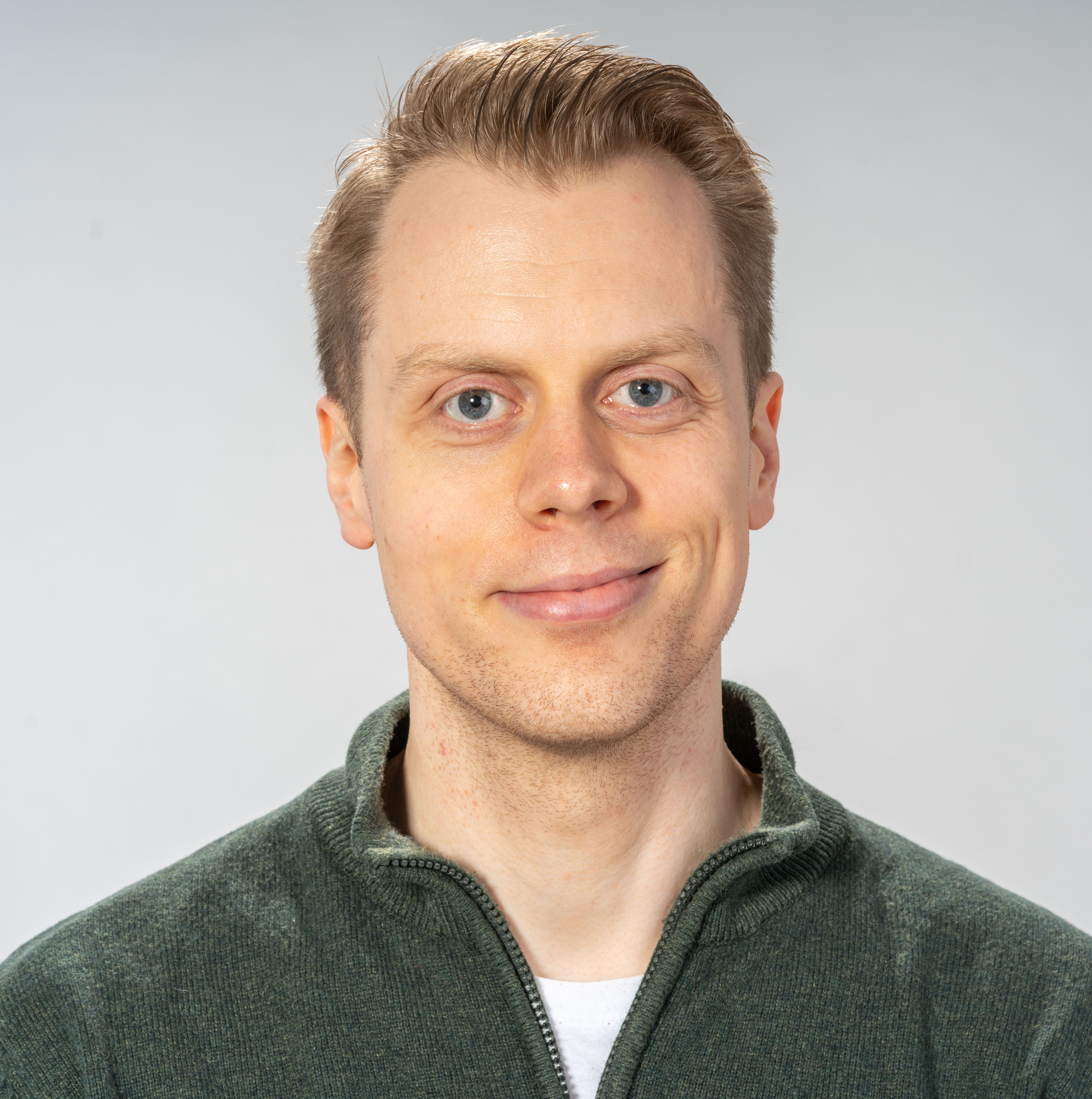}}]{Jørgen Andreas Agersborg}
received the M.Sc.\ degree in Electrical Engineering from UiT The Arctic University of Norway in Tromsø, Norway (2012). From 2012 to 2017 he worked as a researcher in Air and Space Systems at the Norwegian Defence Research Establishment (FFI), Kjeller, Norway. Since 2017 he has been working towards the Ph.D.\ degree in the Machine Learning Group at the Department of Physics and Technology, UiT. His research interests includes developing methods for extracting information from remote sensing data, image processing, and machine learning.
\end{IEEEbiography}

\vskip -2\baselineskip plus -1fil
\begin{IEEEbiography}[{\includegraphics[width=1in,height=1.25in,clip,keepaspectratio]{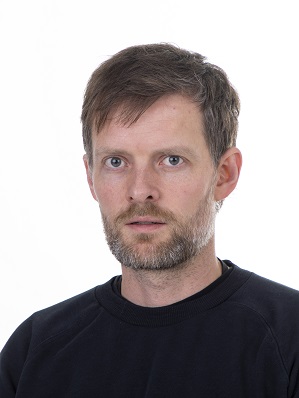}}]{Stian Normann Anfinsen}
received the M.Sc.\ degree in communications, control and digital signal processing from the University of Strathclyde in Glasgow, UK (1998) and the Cand.mag.\ (1997), Cand.scient.\ (2000) and Ph.D.\ degrees (2010) from UiT The Arctic University of Norway in Tromsø, Norway. Since 2014 he has been an Associate Professor with the Department of Physics and Technology at UiT, formerly with the Earth Observation Group and currently as head of the Machine Learning Group. His research interests are in statistical modelling, pattern recognition and machine learning algorithms for image, graph and time series analysis.
\end{IEEEbiography}

\vskip -2\baselineskip plus -1fil
\begin{IEEEbiography}[{\includegraphics[width=1in,height=1.25in,clip,keepaspectratio]{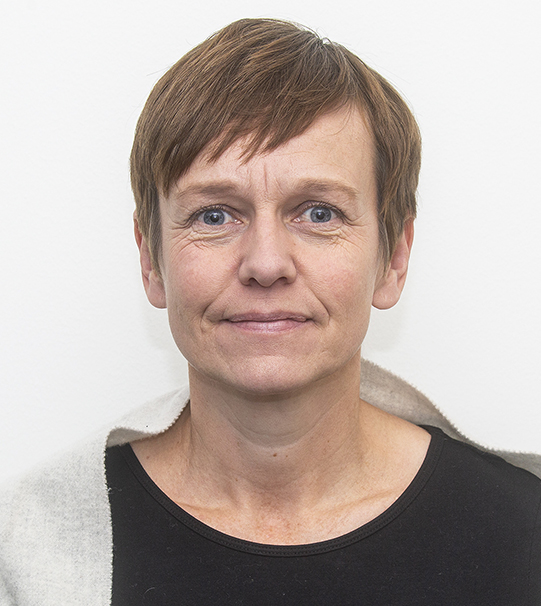}}]{Jane Uhd Jepsen} received the M.Sc.\ degree in biology from the University of Århus, Denmark (1999) and the Ph.D.\ degree in zoology (2004) from University of Copenhagen, Denmark. She is a senior researcher at the Norwegian Institute for Nature Research (NINA) in Tromsø, Norway. Her research interests are in plant-herbivore interactions, population dynamics, and ecological monitoring of northern and Arctic ecosystems. Since 2010 she has been conducting research on the ecosystem consequences of cyclic outbreaks by insect pests in the northern birch forest.  
\end{IEEEbiography}

\end{document}